\documentclass{aa}             %activate for journal version

\input epsf

\begin{document}

% H3954

\title{
Eclipsing binaries in the Magellanic Clouds.
\thanks{Based on observations carried out at the Danish 1.5m telescope at ESO, La Silla, Chile}
}
\subtitle{
$uvby$ CCD light curves and photometric analyses for \\
HV982 (LMC),
HV12578 (LMC),
HV1433 (SMC), and
HV11284 (SMC) 
\thanks{Tables 25-40 will 
be made available on electronic form
at the CDS via anonymous ftp to 130.79.128.5 or via
http://cdsweb.u-strasbg.fr/Abstract.html.
}
}
\author{
J.V. Clausen \inst{1} 
\and J. Storm \inst{2}
\and S.S. Larsen \inst{1,3}
\and A. Gim\'enez \inst{4,5}
}
\offprints{J.V. Clausen}

\institute{
Niels Bohr Institute for Astronomy, Physics and Geophysics,
   Astronomical Observatory,
   Juliane Maries Vej 30, DK-2100 Copenhagen {\O}, Denmark \\
   email: jvc@astro.ku.dk
\and Astrophysikalisches Institut Potsdam,
   An der Sternwarte 16, D-14482 Potsdam, Germany \\
   email: jstorm@aip.de
\and
  UC Observatories / Lick Observatory, University of California, Santa Cruz,
  CA 95064, USA \\
  email: soeren@ucolick.org
\and
Laboratorio de Astrof\'isica Espacial y F\'isica Fundamental
   INTA, Estacion de Villafranca del Castillo,
   Apdo. 50727, E-28080 Madrid, Spain 
\and
Research and Scientific Support Department, ESA, ESTEC, Postbus 299, NL-2200 AG Noordwijk, The Netherlands \\
   email: agimenez@rssd.esa.int
}

\date{received 29.08.2002, accepted 16.12.2002}
 
\titlerunning{$uvby$ CCD light curves}
\authorrunning{J.V. Clausen et al.}

\abstract{
We present new accurate CCD {\it uvby} light curves for the LMC eclipsing binaries
HV982 and HV12578, and for the SMC systems HV1433 and HV11284 obtained at the Danish 
1.5m telescope at ESO, La Silla. 
The light curves were derived from DoPHOT photometry, and typical accuracies are between 
0.007 and 0.012 mag per point. 
Standard {\it uvby} indices have also been established
for each binary, primarily for determination of interstellar reddening and absorption.
For HV982 and HV12578, accurate photometric elements have been established. Both systems
consist of two detached components of comparable sizes in an eccentric orbit.
Adopting the spectroscopic elements given by Fitzpatrick et al. (\cite{fp02}) for HV982,
we derive absolute dimensions of its components which agree well with their results.
A distance modulus of $V_0-M_V = 18.63 \pm 0.08$ is obtained, corresponding to a distance of
$52.6 \pm 2.0$ kpc, which is in formal agreement with (although slightly larger than) 
their determination.
HV1433 and HV11284 both consist of two rather close, deformed and quite different stars.
As the mass ratios between the components (and their rotation rates) are not known,
definitive photometric elements can not yet be obtained, but we present a sample of possible 
photometric solutions.
In a series of forthcoming papers we will combine our {\it uvby} observations with
high-dispersion spectra from the UVES spectrograph on the ESO Very Large Telescope (VLT)
and present absolute dimensions, chemical abundances and distances for
selected LMC and SMC systems, including HV12578 and refined results for HV982.
}

\maketitle

\section{Introduction}
\label{sec:intro}

Double-lined eclipsing binaries provide, apart from objects
where purely geometrical methods can be applied, the most direct
distance determinations available. From complete, accurate observations
and careful analyses of well-detached systems, distance moduli better
than $\pm 0.10$ mag can be obtained. 

With present-day facilities, including ground based 10--m class telescopes 
and the Hubble Space Telescope (HST), precise studies of systems in Local Group 
galaxies, primarily the Magellanic Clouds (MC), have become possible. 
Therefore, accurate \object{LMC} and \object{SMC} 
distances can now be established from eclipsing binaries, leading also to the 
calibrations needed for other important extragalactic di\-stan\-ce indicators, 
like e.g. Cepheids, reaching far beyond the Local Group.
Reviews on the subject have recently been presented by Pacz\'y{n}ski 
(\cite{bp97}) and Clausen (\cite{jvc00}).

Also, the accurate absolute dimensions obtained for the MC binary components 
makes it possible to gain insight into e.g. core structure (e.g Ribas et al \cite{ir00a})
and mass loss for massive stars through comparison with stellar models.
Furthermore, they provide fundamental mass-to-luminosity (M/L) relations for 
metal-deficient stars of e.g. typical LMC/SMC compositions. 

The first accurate CCD light curves of LMC and SMC eclipsing binaries
were obtained at the Danish 1.5m telescope at ESO, La Silla by Jensen et al. 
(\cite{jcg88}), who give references to earlier photometry. The present 
paper should be seen as a continuation of this work. 
Meanwhile, de\-dicated CCD light curve observations have been
done for several LMC and SMC systems at Mt. John University Observatory,
New Zealand (Pritchard et al. \cite{jdpetal98}, \cite{jdp98b} and
references therein), and several thousands of new sy\-stems
have been discovered from the EROS, MACHO, and OGLE
microlensing projects (e.g. Grison et al. \cite{eros95}, Alcock et al.
\cite{macho97}, Udalski et al. \cite{ogle98}).  

Low to medium resolution spectroscopy is available for a few MC systems 
(e.g. Niemela \& Bassino \cite{nb94}), and moderately accurate 
distances and dimensions based on CCD light curves and radial velocities 
from such spectroscopy have been published for some systems, e.g.
\object{HV2241} (Pritchard et al. \cite{jdpetal98}), \object{HV2543} (Ostrov et al.
\cite{o00}), and \object{HV5936} (Bell et al. \cite{sbetal93}) in the LMC, and
\object{HV1620} (Pritchard et al. \cite{jdpetal98}) and \object{HV2226} 
(Bell et al. \cite{sbetal91}) in the SMC.

However, an important step forward was recently taken when a very accurate 
distance modulus (about $\pm 0.07$ mag) 
and stellar dimensions (about $\pm2.5\%$ for radii and $\pm6\%$ for masses) 
were determined for the LMC system \object{HV2274} 
(Watson et al. \cite{w92},
Guinan et al. \cite{g98},
Udalski et al. \cite{u98}
Nelson et al. \cite{n00},
Ribas et al. \cite{ir00b},
Groenewegen \& Salaris \cite{g01}).
Masses and radii were obtained from normal analyses
of light curves and radial velocity curves, based on rather few high 
dispersion spectra, whereas effective temperatures, reddenings and 
distances were obtained from UV/optical spectrophotometry.
Slightly deviating distances are obtained from the
different studies, mainly due to differences in the reddening
determinations, but they all tend to support the short LMC distance scale
(e.g. Gibson \cite{gibson2000}).

Very recently, accurate distances and dimensions for two additional 
LMC systems, HV982 (Pritchard et al. \cite{jdp98b}, Fitzpatrick et al. 
\cite{fp02}) and \object{EROS1044} (Ribas et al. \cite{ir02}), have become available.
A comparison of the results for HV2274, HV982, and EROS1044, and a discussion 
of the derived distances to the optical center of the LMC is given by Ribas et al. 
(\cite{ir02}).
Clearly, studies of many more binaries, well distributed across the LMC, are needed,
and are important also for investigations on the orientation and structure of
the galaxy.
SMC systems should also be included to extend the investigation to significantly 
lower metallicities, which will also allow a study of the metallicity dependence 
of the Cepheid period-luminosity relation. 

In this paper, we present new complete $uvby$ light curves and analyses
of them for the LMC systems HV982 and HV12578, and the SMC system HV11284, 
and partial results for the SMC system HV1433.
Studies of redde\-ning and metallicity in the observed fields (about 
$6.5 \times 6.5$ arc\-mi\-nu\-tes) centered on the four systems have previously 
been published (Larsen \cite{ssl96}, Larsen et al. \cite{lcs01}).

Further, together with a group of colleagues we are undertaking a large scale 
project, combining light curve observations, standard photometry, and high 
dispersion spectroscopy from the ESO VLT UT2 telescope and the new very 
efficient UVES echelle spectrograph (Dekker et al. \cite{dekker00}, 
D'Odorico \cite{dodorico00}). Until now, complete spectroscopic
observations have been obtained for four LMC systems, including HV982 and HV12578,
and two SMC systems, and analyses are in progress.

\section{Selection of candidates}
\label{sec:cand}

\begin{table}
\caption[]{\label{tab:fields}The observed eclipsing binaries. For comparison, the
position of the optical centers of LMC and SMC are included.}
\begin{flushleft}
\begin{tabular}{lrrll} \hline
      ID         &$V_{max}$& $P_{days}$ & $\alpha$(2000.0) & $\delta$(2000.0) \\ \hline
LMC               &        &            & $05^h 24^m$      & $-69\degr45\arcmin$\\
\object{HV982}   & 14.7    &  5.34      & $05^h 29^m 53^s$ & $-69\degr09\arcmin23\arcsec$ \\  % HV 982
\object{HV12578} & 15.8    &  2.48      & $05^h 21^m 32^s$ & $-66\degr21\arcmin15\arcsec$ \\ % HV 12578
SMC               &        &            & $00^h 53^m$      & $-72\degr50\arcmin$\\
\object{HV1433}  & 16.4    &  2.05      & $00^h 47^m 11^s$ & $-73\degr41\arcmin18\arcsec$ \\  % HV 1433
\object{HV11284} & 17.1    &  3.63      & $00^h 49^m 43^s$ & $-72\degr51\arcmin10\arcsec$ \\  % HV 11284
\hline
\end{tabular}
\end{flushleft}
\end{table}
 % fields/binaries, table 1

When the $uvby$ observations were planned, the thousands of newly discovered
MC binaries from the microlensing projects were not yet available,
and our source for the selection was then limited to the about 110 systems
identified mainly from the photographic material of the Harvard surveys 
(Shapley \& Nail \cite{sn42}, \cite{sn53};
Payne-Gaposchkin \& Gaposhkin \cite{pgg66};
Gaposhkin \cite{g70}, \cite{g77};
Hodge \& Wright \cite{hw67}, \cite{hw77};
Kreiner \cite{k72}).

\begin{figure} % figure 1
\epsfxsize=85mm
%\epsfbox{fov982.ps}
\epsfbox{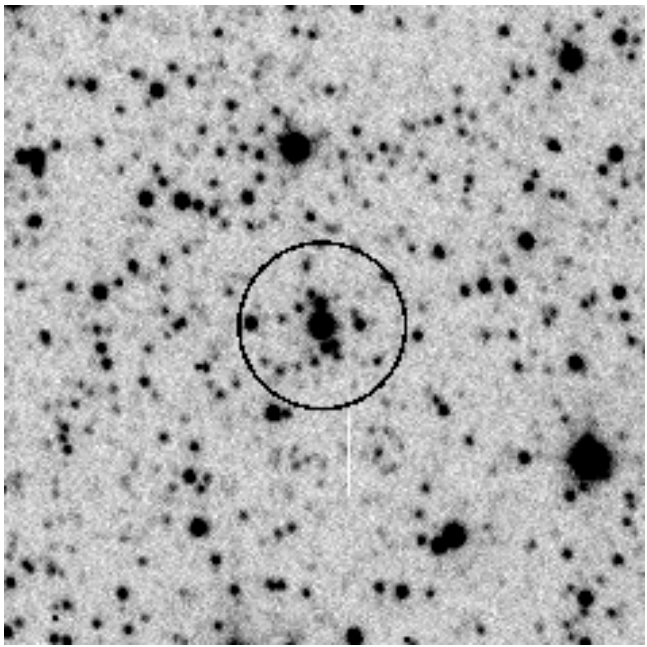}
\caption[]{\label{fig:hv982_field}
HV982 and the surrounding field (1.5 arcmin square). 
The $y$ image was observed at 1.0 arcsec seeing.
North is up, and east is to the left.
}
\end{figure}

\begin{figure} % figure 2
\epsfxsize=85mm
%\epsfbox{fov12578.ps}
\epsfbox{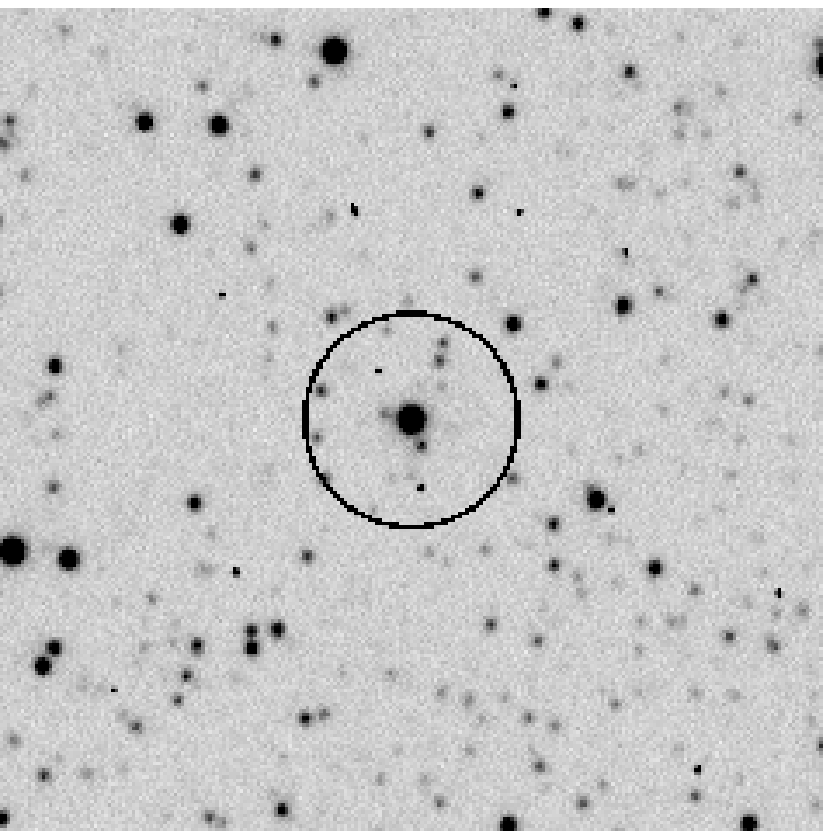}
\caption[]{\label{fig:hv12578_field}
HV12578 and the surrounding field (1.5 arcmin square). 
The $b$ image was observed at 1.1 arcsec seeing.
North is up, and east is to the left.
}
\end{figure}

\begin{figure} % figure 3
\epsfxsize=85mm
%\epsfbox{fov1433.ps}
\epsfbox{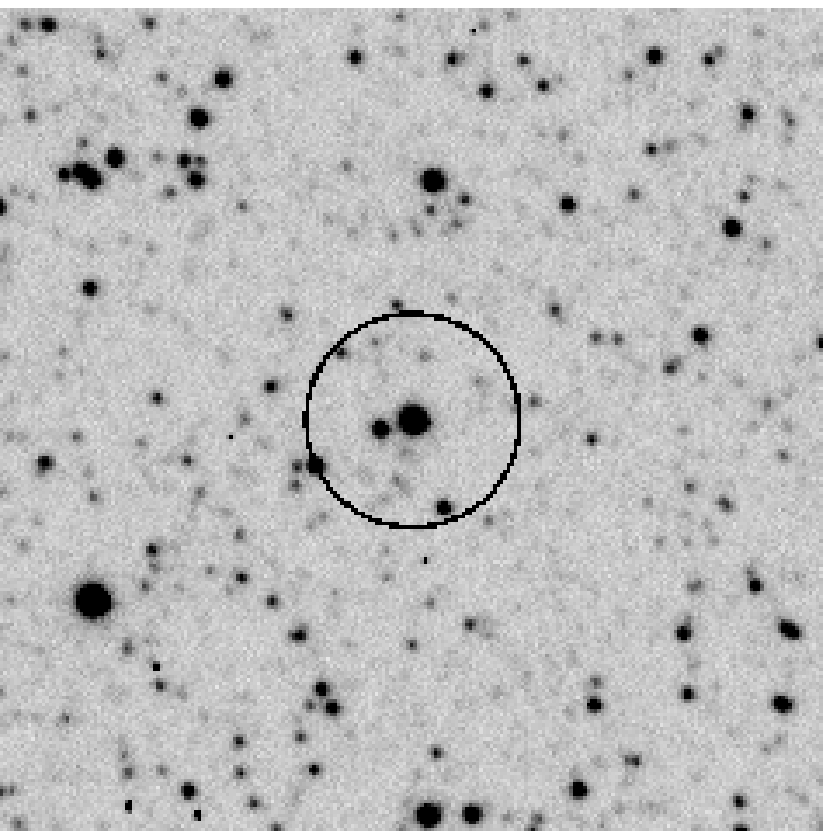}
\caption[]{\label{fig:hv1433_field}
HV1433 and the surrounding field (1.5 arcmin square). 
The $y$ image was observed at 1.1 arcsec seeing.
North is up, and east is to the left.
}
\end{figure}

\begin{figure} % figure 4
\epsfxsize=85mm
%\epsfbox{fov11284.ps}
\epsfbox{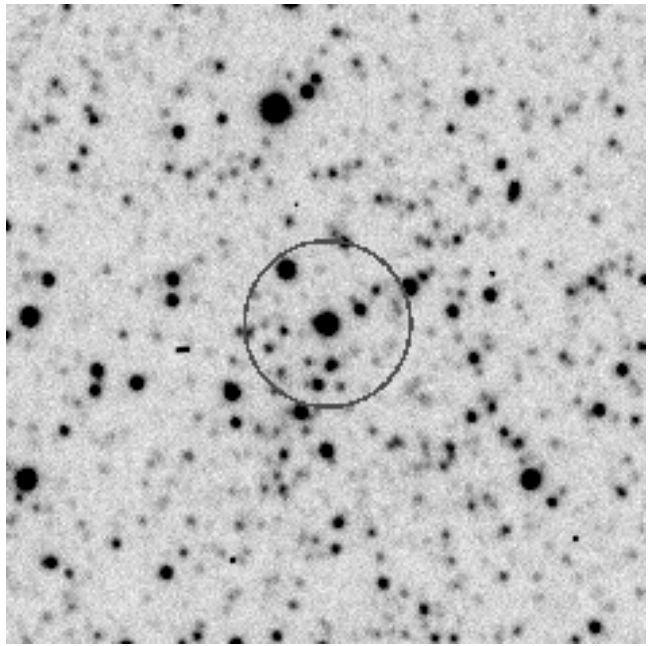}
\caption[]{\label{fig:hv11284_field}
HV11284 and the surrounding field (1.5 arcmin square). 
The $b$ image was observed at 1.1 arcsec seeing.
North is up, and east is to the left.
}
\end{figure}

About 10 candidates per galaxy were selected, but 
for several reasons (field crowding, faintness, difficult orbital
periods, lack of observing time, etc.) the project was gra\-dually
concentrated on the four systems listed in Table~\ref{tab:fields}.
Part of the fields, centered on the eclipsing binaries, are
shown in Figs. 
\ref{fig:hv982_field},
\ref{fig:hv12578_field},
\ref{fig:hv1433_field}, and
\ref{fig:hv11284_field}.

HV982 is located about one degree NE of the optical center of LMC,
whereas HV12578 is located in the northern outskirts of the LMC Bar, 
about three degrees from the optical center.
HV1433 is located in the sou\-thern part of the central region of SMC, just south
of the ${\rm OGLE-II}$ field ${\rm SMC\_SC4}$ (Udalski et al. \cite{ogle98}), and
HV11284 slightly west of the optical center of SMC in the 
${\rm OGLE-II}$ field ${\rm SMC\_SC5}$.
HV11284 is identical to the ${\rm OGLE-II}$ eclipsing binary 
${\rm SMC\_SC5\_140701}$.

\section{Observations and basic reductions}
\label{sec:obsred}

The CCD observations were carried out during several periods 
between November 1992 and November 1995 at the Danish 1.54 meter telescope
at ESO, La Silla, equipped with a direct camera and CCD \#28 (a thinned
$1024 \times 1024$ Tek device). This combination of telescope and CCD
yields a field size of about $6.5 \times 6.5$ arcminutes. 
The image scale is 0.38\arcsec /pixel. The same set of $uvby$
inter\-fe\-ren\-ce filters were used throughout, except for the early 1992 $b$
observations. 
Photometry from several $b$ images of the same field 
obtained with the two filters were carefully compared, but no significant 
differences between the instrumental systems were found. 

The binaries were always positioned near the field centers, but no effort
was made to place them on exactly the same pixels on all exposures.
Also, the orientation of the camera/chip varied between the different 
ob\-ser\-ving runs.
Typical exposure times for HV12578, HV11284, and HV1433 were 5 ($vby$)
and 20 minutes ($u$), but for the brighter HV982 we used 2.5 and 10 minutes,
respectively.  The observations were in general done in $ybvyu$ 
sequences in order to obtain complete light curves in one band ($y$)
and acceptable coverage in the other bands. 

Initial reductions, essentially dark current and bias subtractions and
flatfielding, where carried out using standard
IRAF\footnote{
IRAF is distributed by the National Optical Astronomical Observatories,
which are operated by the Association of Universities for Research in
Astronomy, Inc. under contract with the National Science Foundation.
}
tools. 
As the dark current was very low, only around 4 counts per hour
with no 'hot spots' or other defects visible, we decided to use just a
single common value for the dark current in order not to introduce
additional noise.
In each observing run, typically 10 bias exposures were made, and 
in all cases the bias level was found to be constant within 1-2 ADU
during a given run, whereas the level changed slightly from year to year,
depending on the electronic setup of the CCD. 
The bias frames were all very uniform, the only defect
being a weak pattern from the thinning of the chip, and we used
combined frames for the bias subtraction. 
In each observing run, sky flatfield exposures were made every evening 
and morning, and the individual frames, typically 5 per filter with slight
offsets between them, were carefully checked against a reference frame
obtained with the same filter.
Any exposures that contained bright stars or other obvious anomalies were 
rejected. For the remaining exposures, we found no obvious variations
from night to night or between morning and evening flats, and we thus decided
to use a single combined flatfield in each filter for each run.  The accepted 
flatfield exposures were combined applying a sigma-clipping rejection 
algorithm to avoid any (faint) stars that might be present in the images.

As a last step, we measured the seeing from two or more stars in each frame,
and determined the sky background level. This information is given in
Tables 25-40 (light curve tables, instrumental system). 
No i\-ma\-ges with seeing above about 3 arcsec were accepted, and most 
images were obtained at 1--1.5 arcsec seeing. 
Very few images with a background above 100 ADU were accepted.

Additional information on the reduction is given by Larsen (\cite{ssl96}).

\section{Photometry and formation of light curves}
\label{sec:lc}

\subsection{Light curves}

Several packages are available for crowded field photometry, including 
MOMF (Kjeldsen \& Frandsen \cite{hksrf92}), ROMAPHOT 
(Buonnano et al. \cite{b93}), DoPHOT (Schechter et al.  \cite{schecter93}), 
and DAOPHOT (Stetson \cite{stetson87}). Especially the latter two have
been widely used, and while DAOPHOT may be the most flexible of the two,
DoPHOT is particularly well suited for reductions of
large quantities of data because only a minimum of user interaction is
required. Tests of DoPHOT and DAOPHOT by ourselves and others 
(e.g. Larsen 1996; Ferrarese et al.\ \cite{ferra00}) show that the two packages
generally provide photometry of comparable accuracy, as well as very rea\-li\-stic
error estimates. Because of its large number of user-definable parameters
and greater flexibility, DAOPHOT may give slightly more accurate photometry 
in difficult cases such as strongly crowded fields, or if the PSF varies
significantly across the frame. This may be partly due to the more
sophisticated PSF modeling in DAOPHOT.  However, DoPHOT also produces very 
good photometry and runs substantially faster than DAOPHOT.  Because of the 
large volume of data to be reduced in this project, we therefore opted to
use DoPHOT for the analysis of the LMC and SMC fields.

\begin{table}
\caption[]{\label{tab:dophoterrors}
Mean DoPHOT errors (mag.).
}
\begin{flushleft}
\begin{tabular}{lrrrr} \hline
\hline
Binary      & $y$    &  $b$    &   $v$   &  $u$  \\
\hline
HV982       & 0.008  &  0.007  &  0.008  & 0.008 \\
HV12578     & 0.009  &  0.008  &  0.009  & 0.010  \\
HV1433      & 0.009  &  0.008  &  0.010  & 0.012  \\
HV11284     & 0.008  &  0.007  &  0.008  & 0.009   \\
\hline
\end{tabular}
\end{flushleft}
\end{table}
 % table 2

Following the DoPHOT analysis, the photometry output files were combined
into a single database for each binary field using the DBMAG package developed 
by one of us (JS).  Besides efficient book keeping and photometry
extraction tools, it contains tasks for identification and detection of 
variable stars based on the technique by Welch \& Stetson (\cite{ws93}).
For each binary field and photometric band, a master frame was selected,
and all similar frames were then brought to the zero point of the master. 

As a final check of the procedure, we studied one of the databases,
the $y$ band of the HV11284 field, in further detail. A number of 
isolated, supposedly constant stars spanning the magnitude range of
HV11284 through its orbit were extracted, and their mean magnitude and the corresponding
r.m.s. per observation were calculated. No systematic patterns in the individual
deviations from the means were seen for the stars, and r.m.s. values between 
0.009 and 0.011 mag were obtained. 
For all stars, the r.m.s. values are close to the DoPHOT error 
estimates of the single magnitudes, and we conclude that the frames
have been safely brought to the same zero point. 
Therefore, we have chosen to apply directly the database results for the
binary light curves, rather than basing them on a smaller number of preselected
comparison stars, as is often done.

The light curve observations in the instrumental sy\-stem are presented
in Tables 25-40, which will only be available in electronic form. 
For each light curve point, the internal frame number,
the Heliocentric Julian Date corresponding to the time of mid-exposure, 
the orbital phase,
the magnitude and its error (DoPHOT), the sky background 
level (in ADU), and the seeing (in pixels, the scale is 0.38\arcsec /pixel) 
are given. 
Typical errors per point are close to 0.01 mag (Table \ref{tab:dophoterrors}).

Phases were calculated from the individual ephemerides given below.
When possible, times of minima were calculated from the Kwee \& Van Woerden
(\cite{kvw56}) method. Ephemerides were, if possible, determined from least 
squares fits to new and published times of minima, else by fitting the
observations together applying the method by Lafler \& Kinman (\cite{lk65}).
The resulting light curves are shown in Figs. \ref{fig:hv982}, \ref{fig:hv12578},
\ref{fig:hv1433}, and \ref{fig:hv11284}.

\begin{figure} % figure 5
\epsfxsize=75mm
%\epsfbox{zp982.ps}
\epsfbox{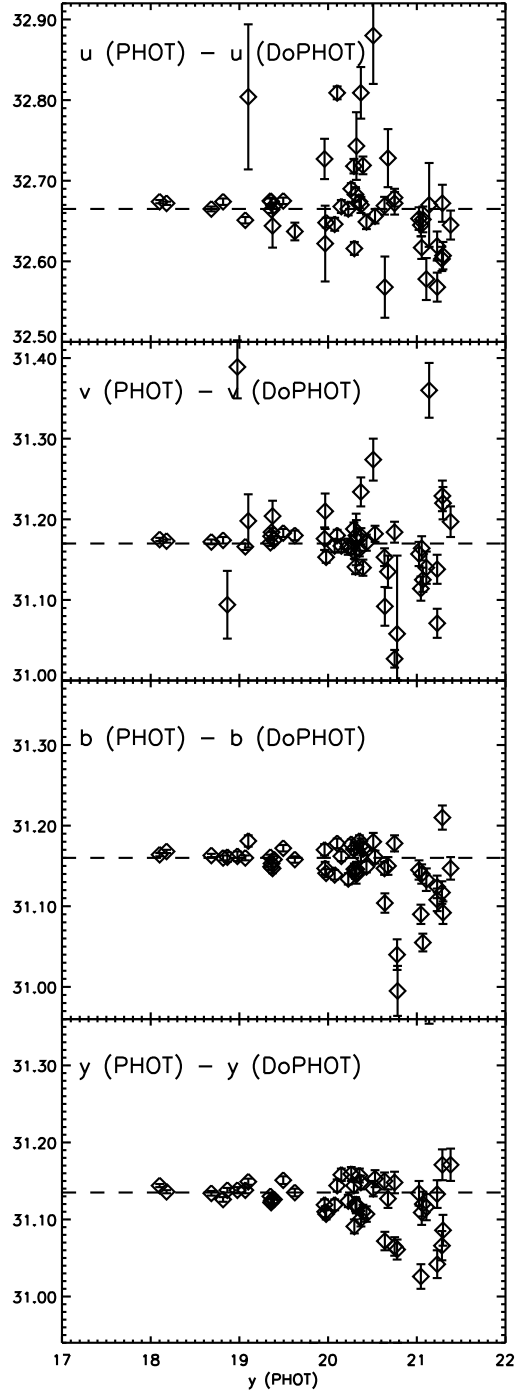}
\caption[]{\label{fig:zp982}Zero-point differences between aperture and 
DoPHOT photometry for ``tertiary standards'' in the HV982 field. The 
dashed lines indicate the adopted zero-points.
}
\end{figure}

\section{Standard $uvby$ indices}
\label{sec:std}

\begin{table*}
\caption[]{\label{tab:std} Standard $uvby$ indices for the eclipsing binaries
(no correction for reddening applied) at maximum light level outside eclipses. 
N is the number of observations used to form the mean value. 
$\sigma$ is the r.m.s. error (per mag/index) given in mmag.}
\begin{flushleft}
\begin{tabular}{lrrrrrrrrrrrr} \hline
\hline
  ID     &$V$&$\sigma$&N&$(b-y)$&$\sigma$&N&$m_1$&$\sigma$&N&$c_1$&$\sigma$&$N$ \\
\hline
HV982    & 14.648&  8& 69&$-0.042$&  9& 47&  0.070 & 15& 44&  0.020 & 17& 42   \\
HV12578  & 15.768&  6& 27&$-0.050$& 10& 20&  0.094 & 13& 16&  0.077 & 25& 11   \\
HV1433   & 15.547& 11&  3&$-0.070$&  5&  4&  0.106 & 14&  4&  0.014 & 25&  3   \\
HV11284  & 15.179&  7& 15&$-0.023$& 10& 27&  0.094 & 15& 21&  0.069 & 13& 17   \\
\hline
\end{tabular}
\end{flushleft}
\end{table*}
 % std indices, table 3
%\input std0.tex % intrinsic std indices, interstellar reddening and absorption, table 4
\begin{table*}
\caption[]{\label{tab:std0} Intrinsic standard $uvby$ indices at maximum light level
outside eclipses, and interstellar reddening and absorption for the eclipsing binaries.}
\begin{flushleft}
\begin{tabular}{lrrrrrrrr} \hline
\hline
  ID    &$ V_0$ & $(b-y)_0$ & $m_0$ & $c_0$ & $E(b-y)$ & $E(B-V)$ & $A_V$ \\
\hline
HV982   & 14.350&   $-0.111$& 0.091 & 0.006 &    0.069 &   0.097  & 0.298 \\
HV12578 & 15.529&   $-0.106$& 0.111 & 0.066 &    0.056 &   0.078  & 0.239 \\
HV1433  & 15.369&   $-0.111$& 0.118 & 0.006 &    0.041 &   0.057  & 0.178 \\
HV11284 & 14.818&   $-0.107$& 0.119 & 0.052 &    0.084 &   0.118  & 0.361 \\
\hline
\end{tabular}
\end{flushleft}
\end{table*}
 % intrinsic std indices, interstellar reddening and absorption, table 4

In order to obtain reliable apparent magnitudes and co\-lors for the
eclipsing binaries, essential for reddening and di\-stan\-ce determinations,
a careful transformation to the $uvby$ standard system is needed.
We have based this calibration on numerous CCD observations
of secondary $uvby$ standard, for which photoelectric observations were
simultaneously done at the Str\"omgren Automatic Telescope (SAT) at ESO, 
La Silla. The SAT observations also provided accurate nightly extinction
coefficient needed for the calibration of the CCD photometry.
The SAT observations and standard $uvby$ indices for the secondary
standards are described by Clausen et al. (\cite{clausen97}). 

Because DBMAG automatically adjusts the photometry of all individual
frames to a common zero-point, it was sufficient to derive a photometric
transformation to the $uvby$ standard system for the reference frames.
For this purpose, we first used the {\bf substar} task in the IRAF DAOPHOT package to subtract 
all stars except a few bright ones from the reference frames. Aperture
photometry for these stars was then obtained with the {\bf phot} task,
thus providing a set of ``tertiary standard stars'' in each
frame (see Larsen et al.\ \cite{lcs01} for details).  Finally, the DoPHOT 
photometry was tied into the standard system by determining the offsets 
between the DoPHOT magnitudes and aperture photometry for the tertiary 
standard stars. Figure~\ref{fig:zp982} shows the difference between the
DoPHOT and aperture photometry for the selected tertiary standards in
the HV 982 field.  Typical uncertainties on the zero-points were estimated 
to be 0.01 -- 0.02 mag.

We further checked the accuracy of the photometric zero-points by comparing 
DAOPHOT PSF-fitting photometry on the \emph{co-added} frames 
(Larsen et al.\ \cite{lcs01}) with the mean DoPHOT 
magnitudes from the DBMAG database. For the LMC fields, the comparison
showed excellent agreement between the DoPHOT and DAOPHOT photometry,
with systematic differences $<0.01$ mag in $y$ and $b-y$ and $<0.02$ mag
in $m_1$ and $c_1$. For the SMC fields we found slightly larger
differences of $\sim0.02$ mag in $y$ and up to 0.03--0.04 mag in
$m_1$, but still compatible with the estimated zero-point uncertainties on 
the photometry. Note that the zero-points were established for the
individual passbands, leading to larger formal uncertainties on indices
such as $m_1$ and $c_1$, which are based on three bands each.

The resulting standard $uvby$ indices for the eclipsing binaries are 
given in Table \ref{tab:std}.  Intrinsic standard $uvby$ indices, interstellar 
reddening, and absorption derived from the $(b-y)_0 - c_0$ relation by 
Crawford (\cite{crawford78}) are given in Table \ref{tab:std0}.  For HV982, 
the reddening determined from the $uvby$ photometry agrees well with that recently 
published by Fitzpatrick et al. (\cite{fp02}), $E(B-V) = 0.086 \pm 0.005$, 
based on a completely different and much more complex approach.

\section{HV982 (LMC)}
\label{sec:hv982}
%HV982 = LMV110
%SAO310, Table 11  2429189.469
%SAO 13, Table 13  2429189.308 5.335268 

\begin{figure*} % figure 6
\epsfxsize=185mm
%\epsfbox{hv982teo.ps}
\epsfbox{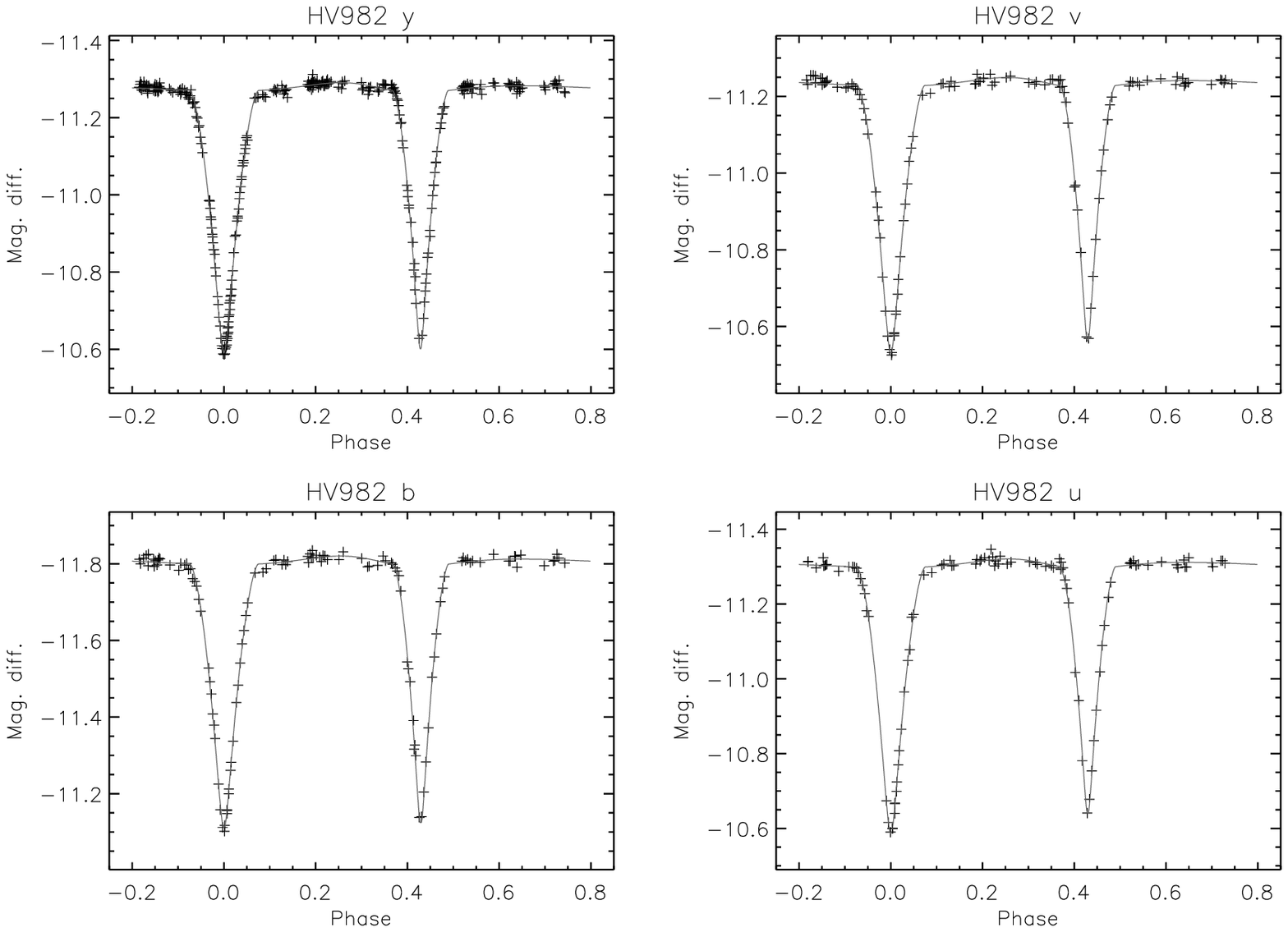}
\caption[]{\label{fig:hv982}
$uvby$ light curves of HV982 (LMC).
The theoretical curves correspond to the solutions given in Table \ref{tab:hv982_wink110}.
}
\end{figure*}

\begin{figure*} % figure 7
\epsfxsize=185mm
%\epsfbox{hv982_2673.ps}
\epsfbox{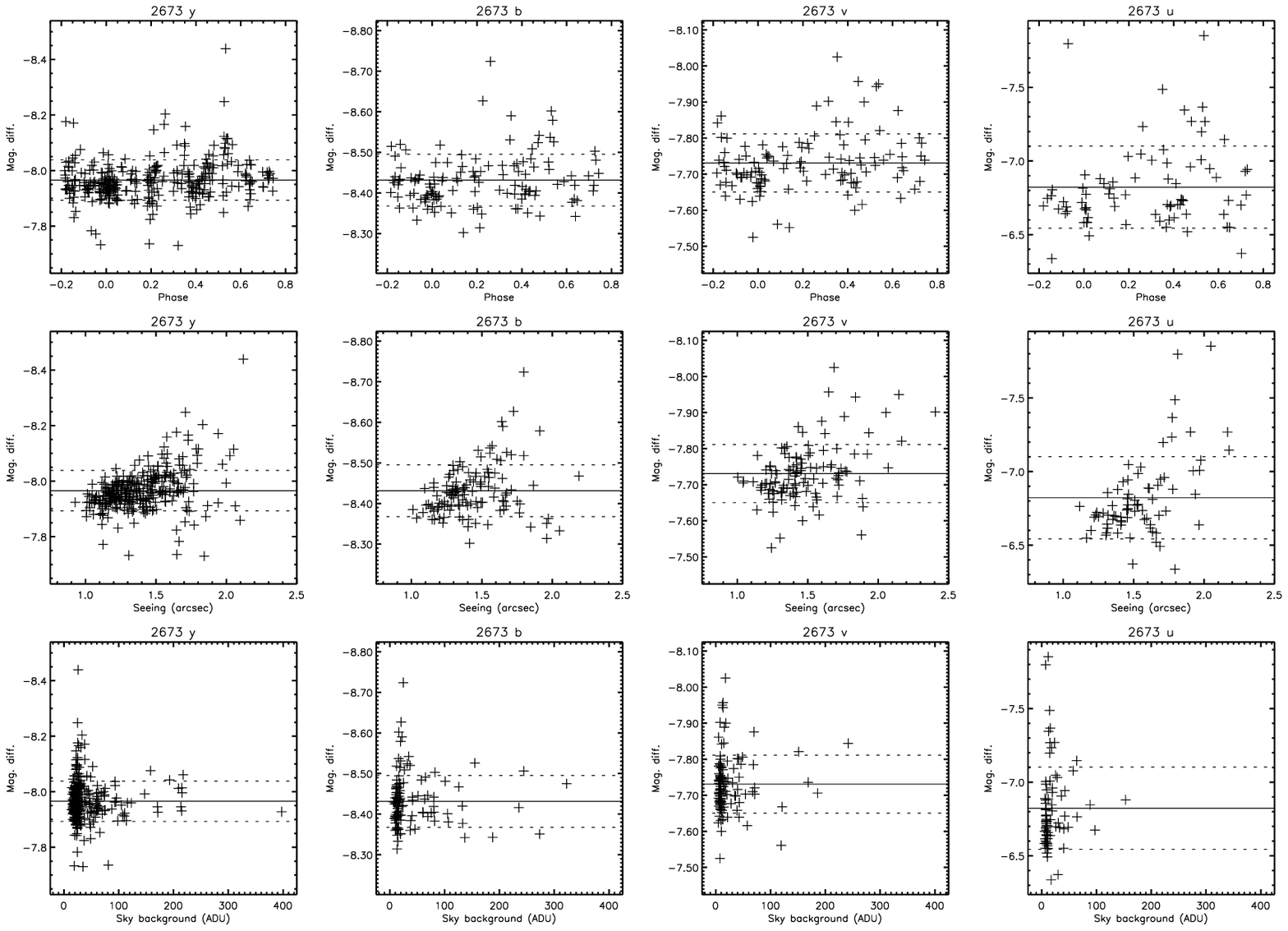}
\caption[]{\label{fig:hv982_2673}
DoPHOT photometry for the close companion (id2673) about 3 arcsec NNE to HV982 (LMC);
see Fig. \ref{fig:hv982_field}.
id2673 is about 3.4 mag fainter than HV982. The full drawn horizontal lines represent the
mean magnitudes, and dashed lines are drawn at the 1 sigma levels.
}
\end{figure*}

Already the photographic Harvard Survey data (Gaposhkin \cite{g70}) show 
that HV982 is a well-detached eclipsing binary, presumably with an 
eccentric orbit. CCD light curves obtained at Mt. John University 
Observatory, New Zealand by Pritchard et al. (\cite{jdp98b}) clearly confirm
these characteristics. Furthermore, with new minima observations, 
apsidal motion with a period of about 200 yr could be detected, making
HV982 just the second MC eclipsing binary with measured orbital motion
(the first known case is HV2274, also in the LMC; Watson et al. \cite{w92},
Ribas et al. \cite{ir00b}).

Pritchard et al. (\cite{jdp98b}) also did standard $uvby$ photometry and 
IUE ultraviolet spectroscopy, and determined interstellar reddening and
photometric elements for HV982. Subsequently, Pritchard \& Niemela
(\cite{pn00}) presented radial velocities and derived mean masses and
radii for the components, and a distance modulus of
$V_0-M_V = 18.7 \pm 0.15$ was obtained. Very recently, Fitzpatrick et al.
(\cite{fp02}) published accurate dimensions for HV982 and found
its distance modulus to be $V_0-M_V = 18.50 \pm 0.06$.
The results were obtained by adding HST UV/optical spectrophotometry
and radial velocities based on high-resolution optical spectroscopy
from the Blanco 4--m  telescope at Cerro Tololo, Chile to existing
photometry for HV982.

As HV982 is clearly an important LMC eclipsing binary, we decided -
parallel to the Mt. John observations - to gain further light curve data.
Mainly because the field near the system is quite complicated, with 
several close neighbours (Fig. \ref{fig:hv982_field}), a delicate subtraction
procedure was needed at the average Mt.\ John seeing of 3--5 arcsec
(Pritchard et al. \cite{jdp98b}). This can be avoided at the a\-ve\-ra\-ge La Silla 
seeing of about 1-2 arcsec. Also, better $uvby$ standard indices were needed. 

In this paper, we present the light curves and photometric elements derived 
from them, plus standard $uvby$ photometry.  Also, absolute dimensions and 
distances are derived, but we expect to publish significantly improved results
when the analyses of the VLT/UVES spectra has been completed.

\subsection{Light curves}
\label{sec:hv982_lc}

HV982 was observed on 43 nights between JD2448970 and JD2450047,
and 314($y$),124 ($b$), 115 ($v$), and 103 ($u$) points were obtained
in the four bands. The light curves are given in Tables 25-28 and
shown in Fig. \ref{fig:hv982}. The coverage in $y$, $b$, and $v$
is fairly complete, whereas a gap exists in the $u$ observations of
primary eclipse. Typical accuracies per point are given in
Table \ref{tab:dophoterrors}.

The two eclipses are of comparable depths with
central secondary eclipse occurring at phase 0.428. 
No significant change of the $(b-y)$ color is seen during the eclipses.
Slight reflection effects are noticed outside the eclipses.

\subsection{The close companions}

As seen on Fig. \ref{fig:hv982_field}, HV982 has several rather close com\-pa\-nions. 
When possible, Pritchard et al. (\cite{jdp98b}) PSF fitted as a group HV982, the two brightest
companions at about 3 arcsec distance (of which the southern actually seems to be two very close stars), 
and a third at about 6 arcsec distance, except at too large seeing where the contributions 
of the companions were subtracted assuming known positions and magnitudes.

As our observations were done at much better seeing, we have been able to
isolate HV982 and avoid significant contamination from all the companions in the 
PSF fits to HV982. This is demonstrated in Fig. \ref{fig:hv982_2673}. 
Here the database photometry for the 3 arcsec distance 
com\-pa\-nion (id2673) NNE to HV982 is plotted versus the orbital phase of HV982 
at the time of the observation, versus the seeing, and versus the sky background level.
The companion, which is the brightest one, is about 3.4 mag ($y$) fainter than HV982 
outside eclipses and its DoPHOT photometry therefore much less accurate.
However, we find no correlation between the magnitude of the companion and the orbital
phase (i.e. light level) of HV982 or the sky background level. There is, perhaps,
a tendency of slightly fainter companion magnitudes for seeing above about 1.6 arcsec
where the scatter on the other hand also increases significantly.

\begin{table}
\caption[]{\label{tab:hv982_tmin}
Times of minima for HV982 (LMC). 
O-C values (in days) are calculated for the apsidal motion parameters given in 
Table \ref{tab:hv982_aps}.
%\ref{eq:hv982_eph}.
References are: G1977, Gaposhkin \cite{g77}; 
P1998, Pritchard et al. \cite{jdp98b}; 
C2003, this paper.
The r.m.s. errors of the G1977 times are estimated.
Note the perfect agreement between P1998 and C2003 for JD2449337.
}
\begin{flushleft}
\begin{tabular}{llcrc} \hline
\hline
HJD-2400000 & r.m.s.   & Type &   O-C   &  Reference  \\    
\hline
13946.555   & 0.050    & P &    0.0290  & G1977  \\ 
17590.584   & 0.050    & P &    0.0350  &  --    \\
23875.527   & 0.050    & P &  $-0.0360$ &  --    \\
25849.645   & 0.050    & P &    0.0130  &  --    \\
26060.243   & 0.100    & S &    0.0900  &  --    \\
26412.253   & 0.050    & S &  $-0.0160$ &  --    \\
26577.631   & 0.050    & S &  $-0.0270$ &  --    \\
27786.315   & 0.050    & P &  $-0.0380$ &  --    \\
29189.469   & 0.100    & P &  $-0.0700$ &  --    \\
29629.338   & 0.050    & S &  $-0.0060$ &  --    \\
31304.630   & 0.050    & S &    0.0520  &  --    \\
32070.603   & 0.050    & P &    0.0002  &  --    \\
33153.625   & 0.050    & P &  $-0.0420$ &  --    \\
49335.3866  & 0.0004   & P &    0.0018  & P1998  \\
49337.6668  & 0.0004   & S &  $-0.0002$ &  --    \\
49337.6670  & 0.0010   & S &    0.0018  & C2003  \\
49340.7172  & 0.0005   & P &  $-0.0028$ & --     \\
50695.8520  & 0.0110   & P &    0.0086  & P1998  \\
\hline
\end{tabular}
\end{flushleft}
\end{table}
 % HV982 times of minima, table 5
%\input hv982_aps.tex % HV982 apsidal motion parameters, table 6
\begin{table}
\caption[]{\label{tab:hv982_aps}
Apsidal motion parameters for HV982 (LMC). 
}
\begin{flushleft}
\begin{tabular}{ll} \hline
\hline
Parameter    & Value and r.m.s. error                    \\
\hline
$i$ (\degr)                & 88.7 (assumed)             \\
$e$                        & 0.159 (assumed)            \\
$T_0$                      & $2449335.17575 \pm 0.00049$ \\
$P_{anomalistic}$ (d)      & $5.335595 \pm 0.000025$     \\ 
$P_{sidereal}$    (d)      &  5.335220                   \\ 
$\omega_0$ (\degr)         & $224.67 \pm 0.15$           \\
$\omega_1$ (\degr / cycle) & $0.00253 \pm 0.00018$       \\
$U$ (yr)                   & $208 \pm 15$                \\
\hline
\end{tabular}
\end{flushleft}
\end{table}
 % HV982 apsidal motion parameters, table 6

\subsection{Ephemeris and apsidal motion}
\label{sec:hv982_eph}

\begin{figure} % figure 8
\epsfxsize=75mm
%\epsfbox{hv982_aps.ps}
\epsfbox{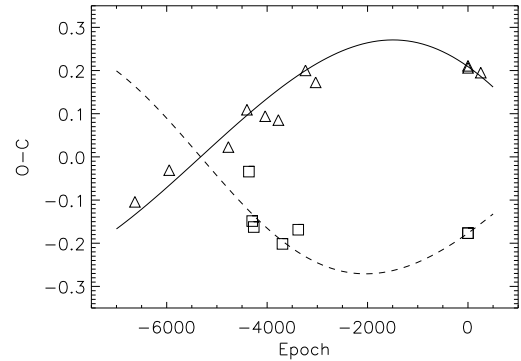}
\caption[]{\label{fig:hv982_aps}
Apsidal motion for HV982 (LMC). O-C are the residuals (days)
from the linear part of the apsidal motion ephemeris defined by the
parameters given in Table \ref{tab:hv982_aps}.
The full and dashed curves represent predictions 
from the apsidal motion parameters for primary (triangles) 
and secondary (squares) eclipses, respectively.
}
\end{figure}

One time of each of primary and secondary minimum, derived
from the $uvby$ CCD observations, are given in Table~\ref{tab:hv982_tmin}
together with times published by Gaposhkin (\cite{g77}) and
Pritchard et al. (\cite{jdp98b}). 

Apsidal motion parameters were derived from a weighted least squares
method, following the formalism by Gim\'enez \& Garcia-Palayo (\cite{ggp83}) 
and Gim\'enez \& Bastero (\cite{gb95}). The orbital inclination $i$ and
eccentricity $e$ were fixed to the values derived from the photometric
analysis; see Sect.~\ref{sec:hv982_lcan}. The results, which are shown
in Table~\ref{tab:hv982_aps} and Fig.~\ref{fig:hv982_aps}, agree quite well 
with those obtained by Pritchard et al. (\cite{jdp98b}).

For the calculation of phases of the $uvby$ light curves, we adopt
the following linear ephemeris derived for the mid-epoch of the 
observations from the apsidal motion parameters:

\begin{equation}
\label{eq:hv982_eph}
\begin{tabular}{r r c r r}
{\rm Min \, I} =  & 2449340.7172 & + & $5\fd 335220$ &$\times\; E$ \\
\end{tabular}
\end{equation}

The adopted period is identical to the sidereal period.
During the light curve observations from JD2448970 to JD2450047, the longitude of 
periastron has changed from $224\fdg5$ to $225\fdg0$ and the phase of mid secondary 
eclipse relative to mid primary eclipse from 0.428 to 0.429. 
For the light curve analyses, the effect of apsidal motion can be ignored, 
except for a few points during secondary eclipse from JD2450041, which
have been excluded.

\subsection{Light curve analysis, photometric elements}
\label{sec:hv982_lcan}

\begin{table}
\caption[]{\label{tab:hv982_winkk} 
WINK solutions ($y$) for HV982 for a range of fixed $k$ values between 0.90 and 1.10. 
$T_p = 24200$ K and $x = 0.33$ have been assumed. 
}
\begin{flushleft}
\begin{tabular}{lrrrrr} \hline
\hline
                   &   $y$   &   $y$   &   $y$   &   $y$   &   $y$\\
\hline
$i (\degr)$        &  88.71  &  88.26  &  88.15  &  88.25  &  88.66\vspace{-0.8mm}\\
                   & $\pm14$ & $\pm10$ & $\pm10$ & $\pm11$ & $\pm16$\\

$e{\rm cos}\omega$ &$-0.1115$&$-0.1115$&$-0.1115$&$-0.1116$&$-0.1117$\vspace{-0.8mm}\\
                   & $\pm  2$& $\pm  2$& $\pm  2$& $\pm  2$& $\pm  2$\\

$e{\rm sin}\omega$ &$-0.1243$&$-0.1194$&$-0.1170$&$-0.1137$&$-0.1124$\vspace{-0.8mm}\\
                   & $\pm 43$& $\pm 43$& $\pm 43$& $\pm 44$& $\pm 44$\\

$e$                & 0.1669  & 0.1634  &  0.1617 &  0.1593 & 0.1585\\

$\omega (\degr)$   & 228.11  & 226.96  &  226.38 &  225.53 & 225.20\\

$k$                &  0.90   &  0.95   &    1.00 &    1.05  &   1.10\\

$r_p$              &  0.2138 &  0.2085 &  0.2033 & 0.1982  & 0.1931\vspace{-0.8mm}\\
                   & $\pm  7$& $\pm  7$& $\pm  7$& $\pm 7$ &$\pm  7$\\

$r_s$              &  0.1924 & 0.1981  & 0.2033  & 0.2081 & 0.2125\\

$r_p+r_s$          &  0.4062 & 0.4066  & 0.4066  & 0.4063 & 0.4056\\

$T_s$  (K)         &  23740  &  23613  &   23572 &  23577 &  23650\vspace{-0.8mm}\\
                   &$\pm103$ &$\pm104$ &$\pm 101$&$\pm104$&$\pm104$\\

$J_s/J_p$          &  0.959  &  0.950  &  0.947  &  0.947 & 0.952\\

$L_s/L_p$          &  0.773 &   0.856  &  0.948  &  1.049 & 1.160\\

$\sigma ({\rm mag}$)& 0.0098&   0.0099 &  0.0098 &  0.0100& 0.0100\\
\hline
\end{tabular}
\end{flushleft}
\end{table}
 % HV982 WINK solutions k=0.90-1.10, table 7
%\input hv982_wink110.tex % HV982 WINK solutions k=1.10, table 8
\begin{table}
\caption[]{\label{tab:hv982_wink110} 
WINK solutions for HV982. $k = r_s/r_p = 1.10$ and $T_p = 24200$ K 
has been assumed. 
}
\begin{flushleft}
\begin{tabular}{lrrrr} \hline
\hline
                   &   $y$   &   $b$   &   $v$   &   $u$   \\
\hline
$i (\degr)$        &  88.66  &  88.96  &  88.84  &  88.80\vspace{-0.8mm}\\
                   & $\pm16$ & $\pm32$ & $\pm26$ & $\pm28$\\

$e{\rm cos}\omega$ &$-0.1117$&$-0.1106$&$-0.1106$&$-0.1103$\vspace{-0.8mm}\\
                   & $\pm  2$& $\pm  4$& $\pm  4$& $\pm  5$\\

$e{\rm sin}\omega$ &$-0.1124$&$-0.1102$&$-0.1065$&$-0.1097$\vspace{-0.8mm}\\
                   & $\pm 44$& $\pm 75$& $\pm 71$& $\pm 76$\\

$e$                & 0.1585  & 0.1561  &  0.1535 &  0.1555\\

$\omega (\degr)$   & 225.20  & 224.90  &  223.91 &  224.83\\

$r_p$              &  0.1931 &  0.1929 &  0.1912 & 0.1932\vspace{-0.8mm}\\
                   & $\pm  7$& $\pm 12$& $\pm 11$& $\pm12$\\

$r_s$              &  0.2125 & 0.2122  & 0.2103  & 0.2125\\

$r_p+r_s$          &  0.4056 & 0.4051  & 0.4015  & 0.4057\\

$x_p = x_s$        &  0.33   & 0.36    & 0.39    & 0.39\\

$T_s$  (K)         &  23650  &  23671  &   23420 &  23468\vspace{-0.8mm}\\
                   &$\pm104$ &$\pm150$ &$\pm 138$&$\pm107$\\

$J_s/J_p$          &  0.953  &  0.954  &  0.931  &  0.907\\

$L_s/L_p$          &  1.160 &   1.160  &  1.134  &  1.105\\

$\sigma ({\rm mag}$)& 0.0100&   0.0104 &  0.0097 &  0.0100\\
\hline
\end{tabular}
\end{flushleft}
\end{table}
 % HV982 WINK solutions k=1.10, table 8

Pritchard et al. (\cite{jdp98b}) presented photometric elements
derived from their CCD light curves, adopting a modified version
of the Wilson-Devinney (WD) code (Wilson \& Devinney \cite{rewd71}; 
Wilson \cite{rew94}). 
Light curves in all bands were a\-na\-ly\-zed simultaneously, 
and four possible sets
of elements were presented, as the differential least squares fit 
did not converge towards one unique solution. 
This reflects a situation which is often seen
for partially eclipsing systems with fairly identical components,
both for circular and eccentric orbits. It is due to significant
correlation between several elements. Often the difference between
theoretical light curves fitted to the observations for a large
range of adopted $k = r_s/r_p$ ($p$ referring to the component
eclipsed at the deeper primary eclipse at phase 0.0) are identical 
within a few mmag.
Pritchard et al. (\cite{jdp98b}) furthermore did not know the
mass ratio $q$ between the components, which is needed for
calculation of the stellar shapes.
The classical ways to break (part of) the degeneracy are to include 
independent spectroscopic information on the luminosity ratio 
(see e.g. Andersen et al. \cite{ja83}), or - for eccentric systems - 
independent information on the orbital eccentricity from apsidal motion 
analyses (see e.g. Clausen et al. \cite{jvc86}).

Fitzpatrick et al. (\cite{fp02}) determined  $q = M_s/M_p = 1.029 \pm 0.027$ 
and the spectroscopic luminosity ratio $L_s/L_p = 1.15 \pm 0.05$ (blue) 
between the components which, 
although the accuracy is not as high as could be desired, enabled them to 
exclude three of the solutions presented by Pritchard et al. (\cite{jdp98b}).

In the following, we will present photometric elements derived from
WINK (Wood \cite{dw71}; extensions from Vaz \cite{lpv84}, \cite{lpv86}; 
Vaz \& Nordlund \cite{lpvan85}; Nordlund \& Vaz \cite{anlpv90})
and WD (Wilson \& Devinney \cite{rewd71}, 
Wilson \cite{rew94}), extensions from Vaz et al. \cite{lpv95})
analyses of the new $uvby$ light curves. Both codes provide physical
binary models, adequate for the description of HV982. They differ
mainly with respect to the model of the stellar shapes
(3-axial ellipsoids with 4th and 5th order terms included versus
Roche potentials) and with respect to the elements determined through 
the differential least squares procedure (e.g. $r_p$ and $k$ versus potentials, 
and $e{\rm cos}\omega$ and $e{\rm sin}\omega$ versus $e$ and $\omega$).
WINK is faster and more used-friendly than WD and therefore well-suited for 
detailed investigations of multi-dimensional parameter spaces. An 
extended IDL-FORTRAN WINK version has been developed by one of us (JVC)
for that purpose.

We have selected to perform independent rather than combined analyses 
of the $uvby$ light curves in order to obtain full understanding of the 
accuracy to which the elements can be established.

\subsubsection{WINK analyses}
\label{sec:hv982_wink}

\begin{figure*} % figure 9
\epsfxsize=185mm
%\epsfbox{hv982_wink.ps}
\epsfbox{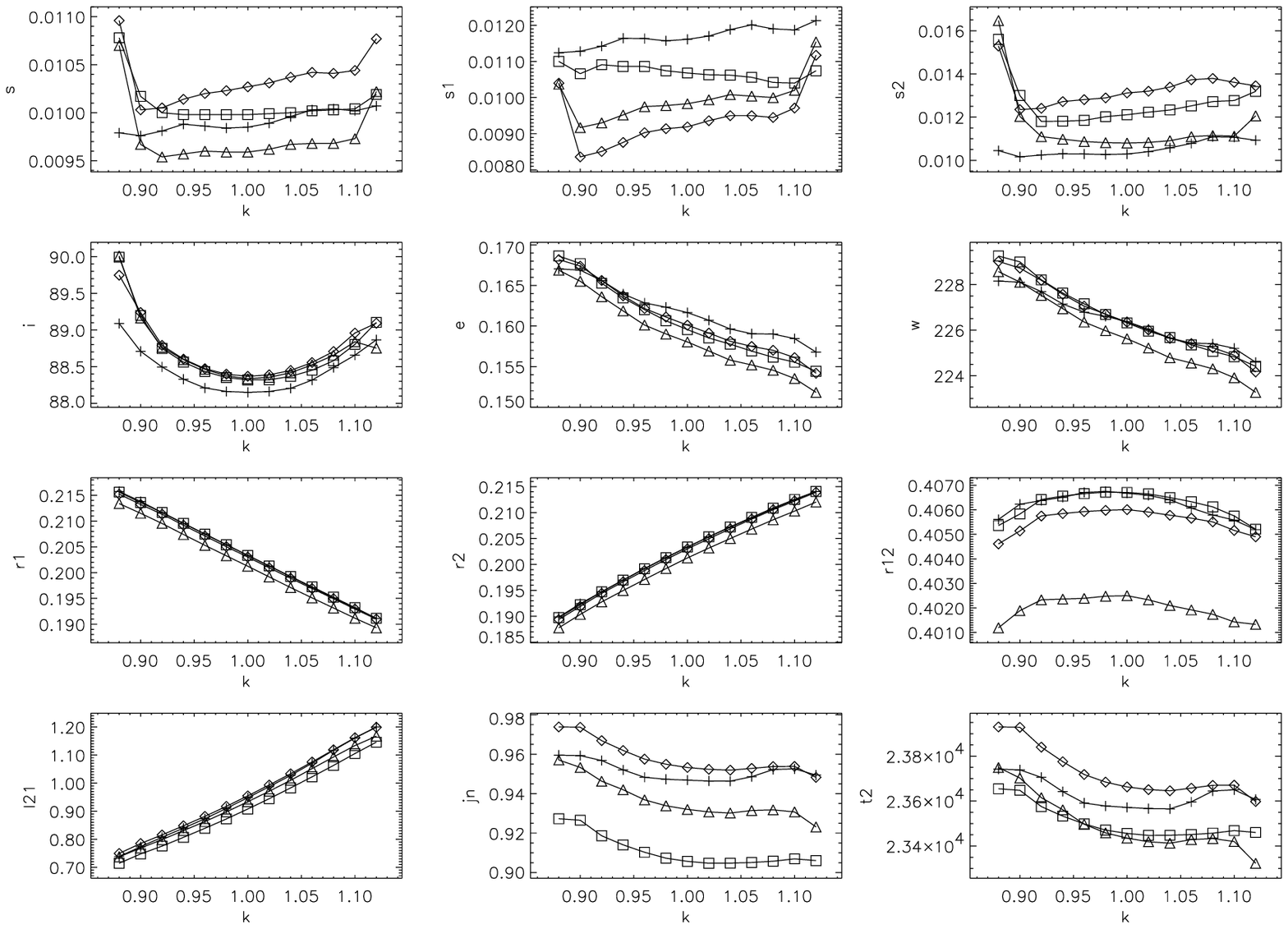}
\caption[]{\label{fig:hv982_wink}
Results from WINK analyses of HV982 for adopted $k = r_s/r_p$ values
between 0.88 and 1.12. The individual figures show
s:  r.m.s. error (mag),
s1: r.m.s. error (mag) for primary eclipse,
s2: r.m.s. error (mag) for secondary eclipse,
i:  orbital inclination $i$, 
e:  orbital eccentricity $e$, 
w:  longitude of periastron $\omega$ ($\degr$),
r1: relative radius for primary component $r_p$,
r2: relative radius for secondary component $r_s$,
r12: sum of relative radii,
l21: luminosity ratio $L_s/L_p$, 
jn:  surface flux ratio,
t2:  effective temperature of secondary component $T_s$ (K).
Symbols are: cross $y$, diamond $b$, triangle $v$, square $u$.
}
\end{figure*}

\begin{figure*} % figure 10
\epsfxsize=185mm
%\epsfbox{hv982_res.ps}
\epsfbox{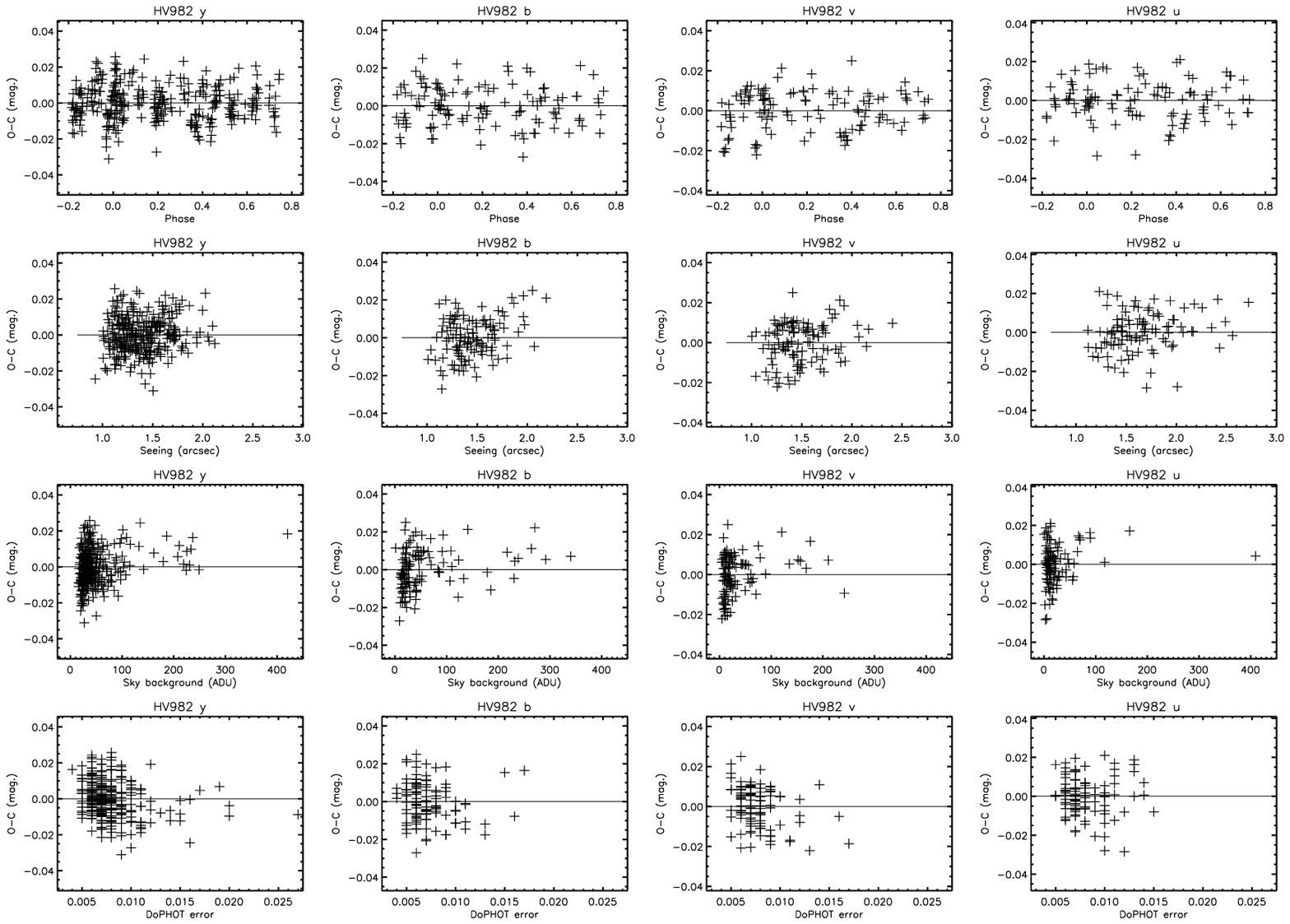}
\caption[]{\label{fig:hv982_res}
$O-C$ residuals for the WINK solutions of HV982 for $k = 1.10$ given in 
Table \ref{tab:hv982_wink110} plotted versus orbital phase, seeing,
sky background, and DoPHOT error (mag).}
\end{figure*}

The results from a series of WINK analyses for fixed $k$ values between
0.88 and 1.12 are shown in Fig. \ref{fig:hv982_wink} and listed in
Table \ref{tab:hv982_winkk}. 
An effective temperature of $T_p = 24200$ K was assumed in agreement
with the combined intrinsic $uvby$ indices given in Table \ref{tab:std0} and 
the calibration by Napiwotzki et al.  (\cite{nap92}), as well as with the energy 
distribution ana\-ly\-sis by Fitzpatrick et al. (\cite{fp02}).
Surface fluxes for the $uvby$ bands, as function of $T_{\mbox{\scriptsize eff}}\,$ 
and ${\rm log}g$ were based on Kurucz (\cite{kurucz92}) ATLAS9 atmosphere models 
(although for Solar composition).
Linear limb darkening coefficients $x$ (see Table \ref{tab:hv982_wink110}) 
by Diaz-Cordoves et al. (\cite{dcg95}) were adopted, and gravity darkening 
exponents of 0.25 and bolometric albedos of 1.0 were assumed in accordance 
with atmospheres in radiative equilibrium. 
For the model calculation of the stellar deformations, a mass ratio of $q = 1.0$ 
was assumed, and the rotation of the stars was 
assumed to be pseudosynchronized (rate at periastron). 
Initially it was settled that no phase shift was needed, i.e. the epoch of 
the ephemeris given in Eq. (\ref{eq:hv982_eph}) is well established.

As seen from the upper panel of Fig. \ref{fig:hv982_wink}, solutions for $k$ 
between 0.90 and 1.10 fit the observations nearly
equally well in all four bands. Several parameters correlate directly
with $k$, primarily $e$ (through $e{\rm sin}\omega$) and $L_s/L_p$,
whereas $T_s$ (and thereby the ratio of surface fluxes $J_s/J_p$) and the
sum of relative radii are nearly constant. Solutions from the
four bands agree quite well - the $y$ solution carries highest weight
due to the larger number of points and the most complete phase coverage for
this band. 
A lu\-mi\-no\-si\-ty ratio ($vby$) of $L_s/L_p = 1.15 \pm 0.05$, as determined by 
Fitzpatrick et al. (\cite{fp02}), corresponds to $k = 1.10 \pm 0.03$. 
We have assumed this value, and solutions for $k=1.10$ are presented 
in Table \ref{tab:hv982_wink110}, and
$(O-C)$ residuals (observed minus calculated mag) are shown in Fig. 
\ref{fig:hv982_res}. No systematic
effects with phase and seeing are seen, but for an unknown reason a 
slight tendency of po\-si\-tive re\-si\-duals (observation fainter than theoretical 
point) for background levels above about 150 ADU seems to be present. 
However, the r.m.s. errors of the fits are comparable to the photometric
errors of the CCD magnitude differences.

\subsubsection{Wilson-Devinney analyses}
\label{sec:hv982_wd}

\begin{table}
\caption[]{\label{tab:hv982_wdy} 
WD  solutions ($y$) for HV982 for fixed $\Omega_p$. 
$T_p = 24200 K$ and $x = 0.33$ have been assumed.
$\Omega$ specify surface potentials and $r$ relative volume radii.
}
\begin{flushleft}
\begin{tabular}{lrrrrr} \hline
\hline
                   &   $y$   &   $y$   &  $y$    &   $y$   &   $y$   \\
\hline
$i (\degr)$        &  88.71  &  88.34  &  88.26  &  88.32  &  88.74\vspace{-0.8mm}\\
                   & $\pm 9$ & $\pm 7$ & $\pm 5$ & $\pm 6$ & $\pm11$\\

$e$                &  0.1633 &  0.1628 &  0.1606 &  0.1593 &  0.1594\vspace{-0.8mm}\\
                   & $\pm 20$& $\pm 21$& $\pm 19$& $\pm 19$& $\pm 22$\\

$\omega$ (\degr)   &  226.76 &  226.58 &  225.80 &  225.34 &  225.37\vspace{-0.8mm}\\
                   & $\pm 67$& $\pm 72$& $\pm 68$& $\pm 69$& $\pm 79$\\

$\Omega_p$         &  5.900  &  6.020  &  6.125  &  6.240  &  6.385 \\

$\Omega_s$         &  6.488  &  6.362  &  6.253  &  6.108  &  6.015\vspace{-0.8mm}\\
                   &$\pm 18$ &$\pm 18$ & $\pm16$ &$\pm 17$ & $\pm16$\\

$r_p$              &  0.2172 & 0.2115  &  0.2066 & 0.2016  & 0.1958\\
$r_s$              &  0.1958 & 0.2007  &  0.2061 & 0.2112  & 0.2155\\

$k$                &  0.902  & 0.949   &  0.998  & 1.048   & 1.101 \\

$r_p+r_s$          &  0.4130 & 0.4122  &  0.4127 & 0.4128  & 0.4113\\

$T_s$ (K)          &  23536  &  23498  &  23471  &   23452 &  23520\vspace{-0.8mm}\\
                   &$\pm 75$ &$\pm 83$ & $\pm73$ &$\pm  65$&$\pm 85$\\

$L_s/L_p$          &  0.778 &   0.859  &  0.947  &  1.044  &  1.157\\

$\sigma ({\rm mag}$)& 0.0097&   0.0097 &  0.0097 &  0.0098 &  0.0099\\
\hline
\end{tabular}
\end{flushleft}
\end{table}
 % HV982 WD solutions (y) for fixed Omega_p, table 9

For comparison, both with the results from our WINK analyses and with the
results by Fitzpatrick et al. (\cite{fp02}), we present in 
Table \ref{tab:hv982_wdy} a series of WD solutions for the $y$ light curve.
The WD solutions were done for very nearly the same ratio of radii 
$k$ as the WINK solutions given in Table \ref{tab:hv982_winkk}, and 
as for the WINK solutions, a mass ratio of $q = 1.0$ and pseudosynchronized
rotation of the components have been assumed. $y$ surface fluxes were
based on Kurucz (\cite{kurucz92}) ATLAS9 atmosphere models, a linear limb darkening law
with $x$ = 0.33, gravity exponents of 1.0 (which correspond to input
values of 0.25 for WINK), and reflection albedos of 1.0, were adopted.

When comparing the results from the two codes, it should first be noted
that in WINK, the relative radii are corrected for the (small)
expansion effects from rotation and gravitational interaction between
the components, whereas such corrections are not applied in the WD
code. In the case of HV982, the corrections are about 0.002 for
both components; see Table \ref{tab:hv982_phel} for further details. 
Taking this code difference into account, the two binary models
are seen to produce practically identical results for HV982. 

\begin{table}            
\caption[]{\label{tab:hv982_phel}
Adopted photometric elements for HV982. $r_p$ and $r_s$ are
corrected for expansion due to rotation and tidal effects, the $r_0$
results are not. The individual flux and luminosity ratios are based
on the mean stellar and orbital parameters.}
\begin{flushleft}             
\begin{tabular}{ll}             
\noalign{\smallskip}             
\hline             
\noalign{\smallskip}             
$i$              & $88{\fdg}7 \pm 0{\fdg}2$ \\
$e\cos \omega$   & $-0.1117 \pm 0.0002$ (at JD 2449500) \\
$e\sin \omega$   & $-0.1124 \pm 0.0044$ (at JD 2449500) \\
$e$              & $0.159 \pm 0.003$ \\             
$\omega$         & $225{\fdg}2 \pm 1{\fdg}1$ (at JD2449500) \\        

$r_p$            & $0.194 \pm 0.003$ \\
$r_s$            & $0.214 \pm 0.003$ \\
$r_p + r_s$      & $0.408 \pm 0.003$ \\
$k$              & $1.10 \pm 0.03$ (assumed)\\
\noalign{\smallskip}             
\end{tabular}             
\begin{tabular}{lrrrrr}             
\noalign{\smallskip}             
                 &    $y$  &    $b$  &    $v$  &    $u$ & \\           
\noalign{\smallskip}             
$J_s/J_p$        & 0.956 & 0.954 & 0.939 & 0.912 & \\ 
$L_s/L_p$        & 1.171 & 1.168 & 1.150 & 1.117 & \\ 
\noalign{\smallskip}             
\end{tabular}             
\begin{tabular}{l}           
\noalign{\smallskip}             
Deformation of the components (WINK model):  \\
\end{tabular}             
\begin{tabular}{lrrrrrr}           
\noalign{\smallskip}
  &  $a$  &   $b$   &   $c$    &  $4th$ ord.  & $5th$ ord. &  $r_0$ \\
Star p    &  0.199  &   0.196 &    0.193    & 0.002  &  0.001 & 0.196
\\ 
Star s    &  0.221  &   0.217 &    0.212    & 0.004  &  0.001 & 0.217
\\
\noalign{\smallskip}            
\hline
\end{tabular}            
\end{flushleft}            
\end{table}  
 % HV982 adopted photometric elements, table 10
%\input hv982_dim.tex % HV982 absolute dimensions, table 11  
\begin{table}   
\caption[]{\label{tab:hv982_dim}
Astrophysical data for HV982.\\
$M_{bol\sun} = 4.74$ has been assumed (Flower \cite{flower96}).
%Last line has r.m.s. from m(0.01), AV(0.02), logR(0.007),
%logTe 0.0045, BC (0.05).
%Maybe we should skip the individual uvby indices, or I should
%at least look at them once more, incl. error estimates.
}
\begin{flushleft}    
\begin{tabular}{lrr}    
\noalign{\smallskip}    
\hline    
\noalign{\smallskip}    
                     &    Primary       &    Secondary      \\ 
\noalign{\smallskip}    
\hline    
\noalign{\smallskip}    
Absolute dimensions:          &                   &                 
 \\ 
$M/M_{\sun}$                  &$11.28 \pm 0.46$   &$11.61 \pm 0.47$ 
\\ 
$R/R_{\sun}$                  & $7.15 \pm 0.12$   & $7.92 \pm 0.13$ 
\\ 
$\log g$ (cgs)                & $3.782 \pm 0.023$ & $3.706 \pm 0.022$
\\ 
 & & \\ 
Photometric data:             &                   &                 
\\ 
$V$     &        15.490   &        15.318 \\
$(b-y)$ &       $-0.044$  &       $-0.041$\\
$m_1$   &         0.062   &         0.077 \\
$c_1$   &         0.012   &         0.027 \\
$E_{(b-y)}$ &     0.068   &         0.082 \\
$A_V$       &     0.293   &         0.309 \\
$(b-y)_0$  &    $-0.112$  &       $-0.111$\\
$c_0$      &    $-0.002$  &         0.013 \\
$[u - b]$ &       0.118   &         0.054 \\
           &              &               \\
$\log T_{\mbox{\scriptsize eff}}\,$ &  $4.384 \pm 0.005$ &   $4.373 \pm 0.005$ \\
$M_{\mbox{\scriptsize bol}}\,$      &  $-5.75 \pm 0.06$ &     $-5.86 \pm 0.07$ \\
$\log L/L_{\sun}$ & $4.19 \pm 0.02$ &    $4.23 \pm 0.03$ \\
$B.C.$            & $-2.31$         &    $-2.25$ \\
$M_V$ &             $-3.44 \pm 0.06$ &   $-3.61 \pm 0.07$ \\
                  &                  &             \\
%$(m-M)_0$          & \multicolumn{2}{c}   {18.63 $\pm$ 0.04} \\
$V_0-M_V$          & \multicolumn{2}{c}   {18.63 $\pm$ 0.08} \\
%Distance \, (kpc) & $53.5 \pm 1.4$   &$53.0 \pm 1.6$ \\
Distance \, (kpc)  & \multicolumn{2}{c}   {52.6 $\pm$ 2.0} \\
\noalign{\smallskip}            
\hline
\end{tabular}            
\end{flushleft}            
\end{table}                                  
 % HV982 absolute dimensions, table 11  

\subsubsection{Adopted photometric elements}
\label{sec:hv982_phel}

From the results of the $uvby$ light curve analyses described above, 
we derive the mean photometric elements for HV982 listed in
Table \ref{tab:hv982_phel}. Within the errors, our results agree
well with those given by Fitzpatrick et al. \cite{fp02}.
Slightly more accurate values have been obtained for the
orbital inclination and eccentricity, probably because our $uvby$
photometry is more accurate than that by Pritchard et al. 
(\cite{jdp98b}). Accepting the luminosity ratio derived by 
Fitzpatrick et al. (\cite{fp02}), both analyses yield
relative radii accurate to about 1\%, i.e. at the level of the
expansion of the components compared to single stars; see Sect.
\ref{sec:hv982_wd}.

\subsection{Absolute dimensions, reddening, and distance}
\label{sec:hv982_disc}

Preliminary absolute dimensions and distance for HV982 are given in 
Table \ref{tab:hv982_dim} together with photometric data for
its components.  The results are based on the 
mean photometric elements listed in Table \ref{tab:hv982_phel} and the 
spectroscopic elements by Fitzpatrick et al. \cite{fp02}.
The $uvby$ indices have been derived from the combined indices at phase
0.25 (Table \ref{tab:std}) and the luminosity ratios (Table \ref{tab:hv982_phel}).
Reddening, interstellar absorption and intrinsic colors are based on
the calibration by Crawford (\cite{crawford78}). 
Effective temperatures were calculated from the empirical [$u-b$] calibration 
by Napiwotzki et al.  (\cite{nap92}), and the well established temperature 
difference from the light curve solution, and bolometric corrections are based on
the calibration by Flower (\cite{flower96}). 

The dimensions and temperatures we obtain for HV982 agree 
well with those presented by Fitzpatrick et al. \cite{fp02}, although 
the photometric data and the methods used for the analyses are different.
The distance has been derived from the 'classical' formula

\begin{equation}
\label{eq:dist}
\begin{tabular}{l l}
$V_0-M_V=$&$V - A_V + 5\times{\rm log}(R/R_\odot) + $\\
          &$10\times{\rm log}(T_{\mbox{\scriptsize eff}}/T_{\mbox{\scriptsize eff}\odot}) + BC - M_{\mbox{\scriptsize bol}\odot}$ 
\end{tabular}
\end{equation}
\noindent
adopting r.m.s. errors of $0.01$ $(V)$, $0.02$ $(A_V)$, and 0.05 $(B.C.)$.
Our value is slightly larger, but in formal agreement with that obtained
by Fitzpatrick et al. \cite{fp02} ($50.2 \pm 1.2$ kpc).
However, we expect to refine both absolute dimensions and distance 
when independent spectroscopic elements, temperatures, luminosity ratios, 
and chemical abundances become available from the VLT/UVES spectra 
mentioned in Sect. \ref{sec:intro}.

Finally, we want briefly to comment on an in\-te\-res\-ting empirical method
presented recently by Salaris and Groenewegen (\cite{sg02}). The method,
which is based on a calibration of Str\"omgren photometry, can be used 
for estimation of distances to B-stars in eclipsing binaries, and among
various systems, they applied the method to HV982.
For both components, a distance of $48.3 \pm5.4$ kpc,
close to the result of the binary analysis, was obtained. 
It is based on the photometry by Pritchard et al. (\cite{jdpetal98}), and the
radii and $V$ data from an earlier version of Fitzpatrick et al. \cite{fp02}, 
which was, however, later revised significantly. 
We have repeated the calculations using the new results listed in 
Table \ref{tab:hv982_dim} and derive distances of
56.5 kpc and 70.9 kpc, respectively, for the components of HV982. 
Such strongly deviating results might indicate that the calibration
by Salaris and Groenewegen needs further refinement.

\section{HV12578 (LMC)}
\label{sec:hv12578}
%HV12578 = LMV1411
%SAO310, Table 11  2429927.447
%SAO 13, Table 13  2429927.447 2.480380 

The photographic light curves of HV12578 (Gaposhkin \cite{g70})
indicate a detached system in a circular orbit with components of 
comparable surface fluxes. The system is relatively bright, 
$m_{pg} = 15.05$ at maximum light, and both eclipses are about 0.7 mag deep.
The orbital period is $2\fd48$, and 
from the widths of the eclipses, relative radii of the components of about 
0.20-0.25 are estimated. Although reflection effects seem to be
present, we judged from the photographic data that HV12578 could very
well be a promising candidates for determination of accurate dimensions
and distances. The field near the binary is not much crowded, but
our first CCD exposures showed that HV12578 has two much fainter 
neighbours at about 3 arcsec distances; see Fig. \ref{fig:hv12578_field}.
At average La Silla seeing they do not cause any problems for the
PSF photometry of HV12578.

\subsection{Light curves}
\label{sec:hv12578_lc}

\begin{figure*} % figure 11
\epsfxsize=185mm
%\epsfbox{hv12578teo.ps}
\epsfbox{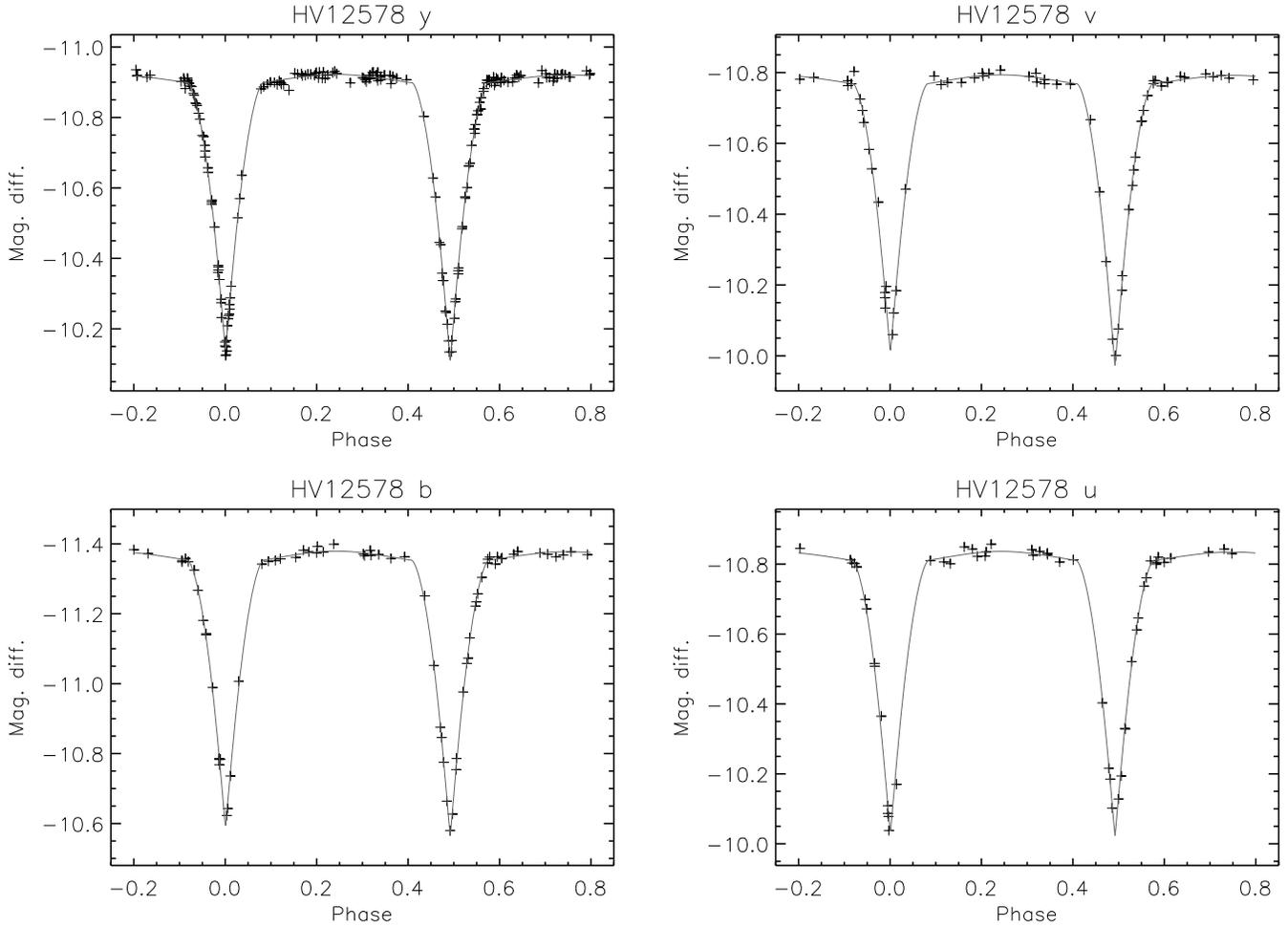}
\caption[]{\label{fig:hv12578}
$uvby$ light curves of HV12578 (LMC).
The theoretical curves correspond to the solutions ($k = 1.00$) given in Table \ref{tab:hv12578_wink}.
}
\end{figure*}

We observed HV12578 on 33 nights between JD2448970 and JD2450040,
and 194 ($y$), 71 ($b$), 65 ($v$), and 53 ($u$) points were obtained
in the four bands. The light curves are given in Tables 29-32 and
shown in Fig. \ref{fig:hv12578}. 
Although the phase coverage is not as complete as we would prefer,
the $y$ and $b$ light curves are adequate for photometric ana\-lyses.
The $v$ and $u$ data are more scarce and, besides providing luminosity
ratios between the components in these bands, they mainly serve as check data.
Typical accuracies per point are given in Table \ref{tab:dophoterrors}.

The new CCD observations reveal that reflection effects are 
smaller than indicated by the Harvard material, and also that secondary
eclipse is shifted slightly from phase 0.50, i.e. the orbit is eccentric.
Both eclipses are deep, about 0.8 mag in all four bands, so the orbital
inclination must be very close to $90\degr$
(a similar galactic F-type case is DM Vir; see Andersen et al. \cite{acn84}).
From the CCD observations alone,
it is not possible to settle definitively which eclipse is the deeper one.
We have therefore decided to keep the $T_0$ definition given by
the Harvard ephemeris (Payne-Gaposchkin \cite{pg71}), although the
final photometric analysis presented in Sect. \ref{sec:hv12578_lcan} 
shows that for this definition, the component eclipsed near phase 0.49 is 
marginally hotter than the other. We will return to this matter once the
mass ratio becomes available from the analysis of the VLT/UVES spectra.

\subsection{Ephemeris}
\label{sec:hv12578_eph}

\begin{table}
\caption[]{\label{tab:hv12578_tmin}
Times of minima for HV12578 (LMC). 
Phases  are calculated from the ephemeris given in Eq. \ref{eq:hv12578_eph}.
References are: G1977, Gaposhkin \cite{g77};
C2003, this paper.
The r.m.s. errors of the G1977 times are estimated.
The minimum at JD2424501.614 listed by G1977 has been omitted.
}
\begin{flushleft}
\begin{tabular}{llcrc} \hline
\hline
HJD-2400000  & r.m.s.    & Type &   Phase    &  Reference  \\    
\hline
16823.638  &   0.050 &P&  0.010 &  G1977 \\
23681.879  &   0.050 &P&  0.010 &  --\\
23907.499  &   0.050 &P&$-0.028$&  --\\
%24501.614  &        & &  &     & -- \\ %   very large residual, not used
26303.577  &   0.050 &P&$-0.015$&  --\\
26328.446  &   0.050 &P&  0.012 &  --\\
27800.507  &   0.050 &S&  0.494 &  --\\
29881.575  &   0.050 &S&  0.507 &  --\\
29927.447  &   0.050 &P&  0.001 &  --\\
29994.482  &   0.050 &P&  0.027 &  --\\
30640.598  &   0.050 &S&  0.518 &  --\\
31496.277  &   0.050 &S&  0.498 &  --\\
48971.777  &   0.001 &P&$-0.000$& C2003 \\
49343.835  &   0.001 &P&  0.000 & --    \\
49637.738  &   0.001 &S&  0.492 & --    \\
50040.825  &   0.010 &P&  0.002 & --    \\
\hline
\end{tabular}
\end{flushleft}
\end{table}
    % HV12578 times of minima, table 12
\begin{table*}
\caption[]{\label{tab:hv12578_wink} 
WINK solutions for HV12578 for $k = r_s/r_p$ values of 0.98, 1.00, and 1.02. 
$i = 90\fdg0$ and $T_p = 22000$ K were adopted. For the $b$, $v$, and $u$ solutions,
$e{\rm cos}\omega$ and $e{\rm sin}\omega$ were fixed at the $y$ results.
}
\tiny{
\begin{flushleft}
\begin{tabular}{lrrrrrrrrrrrr} \hline
\hline
                    &   $y$   &   $b$   &   $v$   &   $u$   &   $y$   &   $b$   &   $v$   &   $u$   &   $y$   &   $b$   &   $v$   &   $u$\\
\hline
$k$                 &   0.98  &   0.98  &   0.98  &   0.98  &   1.00  &  1.00   &  1.00   &  1.00   &   1.02  &   1.02  &   1.02  &   1.02\\

$e{\rm sin}\omega$  & 0.00648 & 0.00648 & 0.00648 & 0.00648 & 0.00745 & 0.00745 & 0.00745 & 0.00745 & 0.01076 & 0.01076 & 0.01076 & 0.01076\vspace{-0.6mm}\\  
                    &$\pm 551$&         &         &         &$\pm 546$&         &         &         &$\pm 555$&         &         &        \\

$e{\rm cos}\omega$  &-0.01341&-0.01341&-0.01341&-0.01341&-0.01341&-0.01341&-0.01341&-0.01341&-0.01341&-0.01341&-0.01341&-0.01341\vspace{-0.6mm}\\
                    &$\pm  31$&         &         &         &$\pm  30$&         &         &         &$\pm  31$&         &         &        \\

$e$                 &  0.0149 &  0.0149 & 0.0149  & 0.0149  & 0.0153  & 0.0153  &  0.0153 &  0.0153 &  0.0172 & 0.0172  & 0.0172  & 0.0172  \\

$\omega (\degr)$    &  154.2  &  154.2  &  154.2  &  154.2  & 150.9   & 150.9   & 150.9   &  150.9  &  141.3  & 141.3   & 141.3   &  141.3  \\

$r_p$               &  0.2450 &  0.2479 &  0.2442 &  0.2497 &  0.2426 &  0.2455 &  0.2419 & 0.2472  &  0.2402 &  0.2430 &  0.2395 &  0.2447\vspace{-0.6mm}\\
                    & $\pm 10$& $\pm 13$& $\pm 2$£& $\pm 25$& $\pm 10$& $\pm 14$& $\pm 20$& $\pm23$ & $\pm 10$& $\pm 13$& $\pm 22$& $\pm 23$\\

$r_s$               & 0.2401  &  0.2429 &  0.2394 &  0.2447 &  0.2426 &  0.2455 &  0.2419 & 0.2472  &  0.2450 &  0.2478 &  0.2443 &  0.2496  \\

$r_p+r_s$           & 0.4851  &  0.4908 &  0.4836 &  0.4944 &  0.4852 &  0.4910 &  0.4838 & 0.4944  &  0.4852 &  0.4908 &  0.4838 &  0.4943\\

$x_p = x_s$         & 0.32    &  0.37   &  0.39   &  0.39   &  0.32   & 0.37    & 0.39    & 0.39    & 0.32    &  0.37   &  0.39   &  0.39\\

$T_s$ (K)           &  22328  &  22328  &  22447  &  22079  &  22332  &  22295  &   22430 &  22073  &  22277  &  22289  &  22409  &  22058\vspace{-0.6mm}\\
                    &$\pm 76$ &$\pm100$ &$\pm151$ &$\pm107$ &$\pm 75$ &$\pm100$ &$\pm 151$&$\pm108$ &$\pm 76$ &$\pm100$ &$\pm151$ &$\pm108$ \\

$J_s/J_p$           &  1.027  &  1.028  &  1.041  &  1.010  &  1.027  &  1.025  &  1.039  &  1.009  &  1.023  &  1.024  &  1.037  &  1.007\\

$L_s/L_p$           &  0.986  &  0.987  &  0.999  &  0.971  &  1.028 &   1.026  &  1.040  &  1.011  &  1.066  &  1.068  &  1.081  &  1.051\\

$\sigma ({\rm mag}$)&  0.0109 &  0.0093 &  0.0148 &  0.0157 & 0.0107&   0.0094 &  0.0149 &  0.0159  &  0.0111 &  0.0093 &  0.0149 &  0.0160\\
\hline
\end{tabular}
\end{flushleft}
}
\end{table*}
    % HV12578 WINK solutions, table 13  
\begin{table}
\caption[]{\label{tab:hv12578_wdy} 
WD  solutions ($y$) for HV12578 for fixed $\Omega_p$. \\
$i = 90\fdg0$, $T_p = 22000 K$ and $x = 0.32$ have been assumed.
}
\begin{flushleft}
\begin{tabular}{lrrr} \hline
\hline
                   &   $y$   &   $y$   &  $y$       \\
\hline
$e$                &  0.0157 &  0.0153 &  0.0171\vspace{-0.8mm}\\
                   & $\pm 12$& $\pm 14$& $\pm 20$\\

$\omega (\degr)$   &  150.7  &  150.3  &  139.4\vspace{-0.8mm}\\
                   & $\pm6.9$& $\pm8.0$& $\pm6.9$\\

$\Omega_p$         &  5.040  &  5.120  &  5.200\\

$\Omega_s$         &  5.117  &  5.126  &  5.138\vspace{-0.8mm}\\
                   &$\pm 18$ &$\pm 16$ & $\pm17$\\

$r_p$              &  0.2513 & 0.2462  &  0.2414\\
$r_s$              &  0.2464 & 0.2458  &  0.2452\\

$k$                &  0.981  & 0.998   &  1.016\\

$r_p+r_s$          &  0.4977 & 0.4920  &  0.4866\\

$T_s$  (K)         &  22502  &  22512  &  22461\vspace{-0.8mm}\\
                   &$\pm 54$ &$\pm 57$ & $\pm57$\\

$L_s/L_p$          &  0.983 &   1.020  &  1.052\\

$\sigma ({\rm mag}$)& 0.0115&   0.0114 &  0.0121\\
\hline
\end{tabular}
\end{flushleft}
\end{table}
    % HV12578 WD solutions (y), table 14
\begin{table}            
\caption[]{\label{tab:hv12578_phel}
Adopted photometric elements for HV12578. $r_p$ and $r_s$ are
corrected for expansion due to rotation and tidal effects, the $r_0$
result (both components) is not. 
The individual flux and luminosity ratios are based
on the mean stellar and orbital parameters.}
\begin{flushleft}             
\begin{tabular}{ll}             
\noalign{\smallskip}             
\hline             
\noalign{\smallskip}             
$i$              & $90{\fdg}0 \pm 0{\fdg}2$ \\
$e\cos \omega$   & $-0.01341 \pm 0.00030$  \\   
$e\sin \omega$   & $ 0.00745 \pm 0.00200$  \\    % .0065 to .0108
$e$              & $0.0153 \pm 0.0010$ \\             
$\omega$         & $151{\fdg}0 \pm 6{\fdg}5$ \\      %141.3 154.2

$r_p$            & $0.2425 \pm 0.0025$ \\
$r_s$            & $0.2425 \pm 0.0025$ \\
$r_p + r_s$      & $0.4850 \pm 0.0035$ \\
$k$              & $1.00 \pm 0.02$ \\
\noalign{\smallskip}             
\end{tabular}             
\begin{tabular}{lrrrrr}             
\noalign{\smallskip}             
                 &    $y$  &    $b$  &    $v$  &    $u$ & \\           
\noalign{\smallskip}             
$J_s/J_p$        & 1.027 & 1.026 & 1.039 & 1.011 &   % v,u not very good
\\ 
$L_s/L_p$        & 1.028 & 1.027 & 1.040 & 1.013 &   % v,u not very good
\\ 
\noalign{\smallskip}             
\end{tabular}             
\begin{tabular}{l}           
\noalign{\smallskip}             
Deformation of the components (both; WINK model):  \\
\end{tabular}             
\begin{tabular}{lrrrrrr}           
\noalign{\smallskip}
  &  $a$  &   $b$    &   $c$    &  $4th$ ord.  & $5th$ ord. &  $r_0$ \\
          &  0.2498  &   0.2442 &    0.2408    & 0.0037  &  0.0009 & 0.2449
\\ 
\noalign{\smallskip}            
\hline
\end{tabular}            
\end{flushleft}            
\end{table}                                  
    % HV12578 adopted photometric elements, table 15

Three times of primary minimum (at phase 0.00) and one of secondary minimum
(near phase 0.49), derived
from the $uvby$ CCD observations, are given in Table~\ref{tab:hv12578_tmin}
together with times published by Gaposhkin (\cite{g70}).
Weighted (weight proportional to square of inverse r.m.s.) least squares fits to
the times of primary and secondary minima yield periods of
$2\fd 4803769 \pm 0.0000017$ and
$2\fd 4803730 \pm 0.0000017$,
respectively, and nearly identical results are obtained if equal weights are
applied. For the calculation of phases of the $uvby$ observations, we adopt

\begin{equation}
\label{eq:hv12578_eph}
\begin{tabular}{r r c r r}
{\rm Min \, I} =  & 2448971.7778 & + & $2\fd 4803769$ &$\times\; E$ \\
                  &      $\pm  6$&   &       $\pm 17$ &             \\
\end{tabular}
\end{equation}

The orbital period agrees well with that published by
Payne-Gaposchkin (\cite{pg71}), $2.48038$, but is much more accurate.
The time of secondary eclipse determined from the $uvby$ observations
reveals that the orbit of HV12578 is slightly eccentric, and this is
confirmed by the light curve analysis; see Sect. \ref{sec:hv12578_lcan}. 
The slight difference in the pe\-ri\-ods derived from times of 
primary and secondary eclipses, respectively, could in principle be due to 
apsidal motion, but since the orbital eccentricity is just about 0.015 this
is not very likely, and it can not at all be confirmed from an analysis of
the available times of minima. 

\subsection{Light curve analysis, photometric elements}
\label{sec:hv12578_lcan}

For the analyses of the CCD light curves of HV12578, we have adopted the 
WINK and Wilson-Devinney models described in Sect. \ref{sec:hv982_lcan}.

The combined intrinsic $uvby$ indices given in Table \ref{tab:std0} and the 
empirical [$u-b$] calibration by Napiwotzki et al. (\cite{nap92}) yield a
mean $T_{\mbox{\scriptsize eff}}\,$ of 22000 K, and since the components have nearly identical
surface fluxes, this value has been assigned to the 
primary component (eclipsed at phase 0.00) and kept fixed throughout the analysis.
Linear limb darkening coefficients $x$ (see Table \ref{tab:hv12578_wink})
by Diaz-Cordoves et al. (\cite{dcg95}) were adopted, and gravity darkening
exponents of 0.25 (WINK) and 1.00 (WD), and bolometric albedos of 1.0,
were assumed in accordance with atmospheres in radiative equilibrium.
For the model calculation of the stellar deformations, a mass ratio of 
$q = 1.0$ was assumed, and the rotation of the stars was
assumed to be pseudosynchronized (rate at periastron).

\begin{figure*} % figure 12
\epsfxsize=185mm
%\epsfbox{hv12578_ywink.ps}
\epsfbox{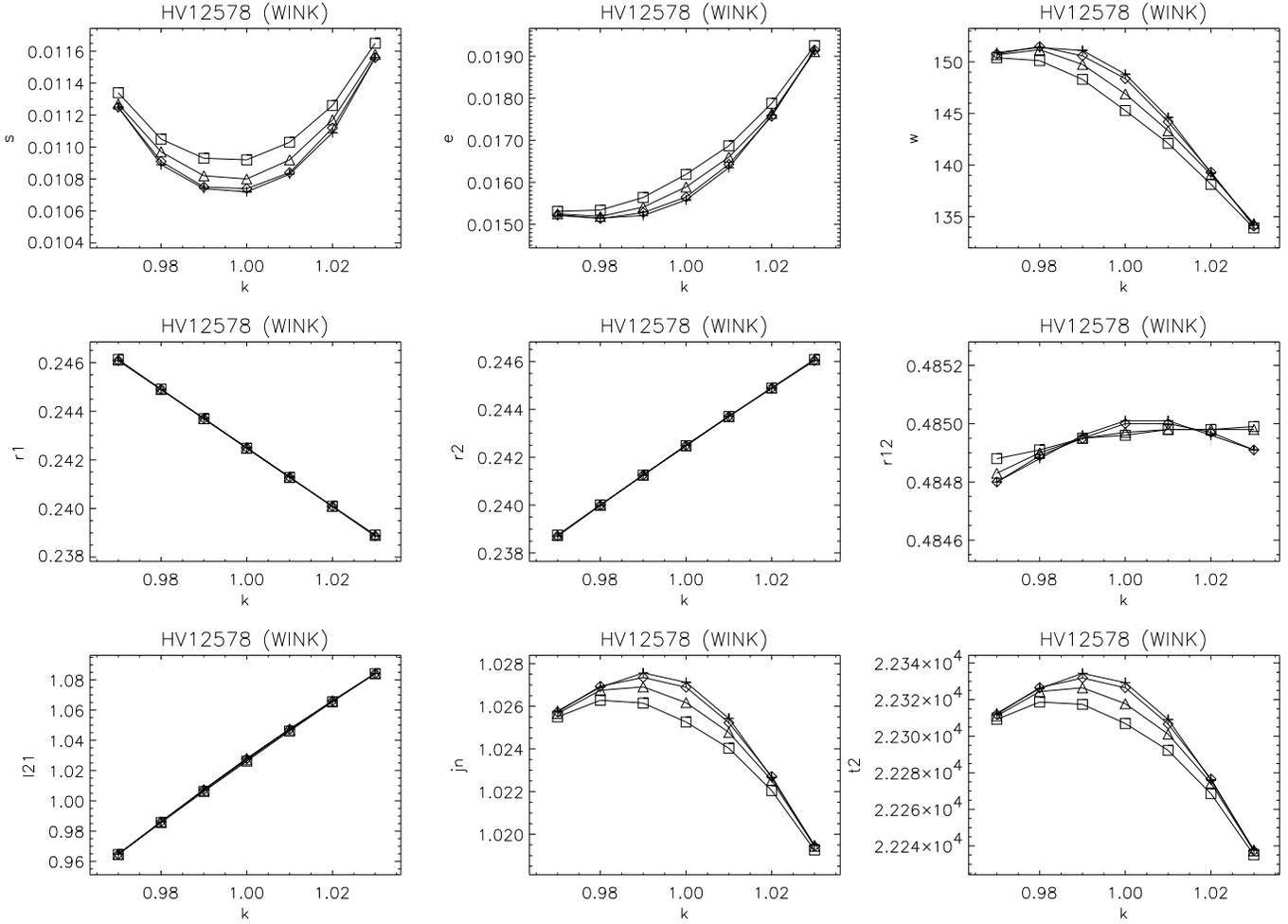}
\caption[]{\label{fig:hv12578_ywink}
Results from WINK analyses of the $y$ light curve of HV12578 
for adopted  $i$ values between $89\fdg7$ and $90\fdg0$, and 
$k = r_s/r_p$ values between 0.97 and 1.03. 
A constant phase shift of 0.00065 has been used.
The individual figures show
s:  r.m.s. error (mag),
e:  orbital eccentricity $e$, 
w:  longitude of periastron $\omega$ ($\degr$),
r1: relative radius for primary component $r_p$,
r2: relative radius for secondary component $r_s$,
r12: sum of relative radii,
l21: luminosity ratio $L_s/L_p$, 
jn:  surface flux ratio,
t2:  effective temperature of secondary component $T_s$ (K).
Symbols are: 
cross $i=90\fdg0$, 
diamond $i=89\fdg9$, 
triangle $i=89\fdg8$, 
square $i=89\fdg7$.
}
\end{figure*}

\subsubsection{WINK analyses}
\label{sec:hv12578_wink}

Normally, at least $i$, $e{\rm cos}\omega$, $e{\rm sin}\omega$, $r_p$,
$T_s$, and possibly $k$, would be kept free and determined through the differential least
squares procedure, but initial tests showed that this approach is not possible for
HV12578, primarily because $i$ is very close to $90\degr$. We have therefore
calculated grids of solutions for adopted values of $i$, each of them for a range of
adopted $k$ values centered at 1.0. 
Fig. \ref{fig:hv12578_ywink} illustrates the results from such WINK solutions 
for the $y$ light curve. It is seen that the best fits are obtained for
$i$ between $89\fdg8$ and $90\fdg0$ and $k$ between 0.98 and 1.02.
For a given $k$, most other parameters - especially the individual radii -
are very little dependent on $i$ within the range mentioned above.
Similar analyses for the other three bands support these findings, but are
less accurate due to the fewer observations.

\begin{figure*} % figure 13
\epsfxsize=185mm
%\epsfbox{hv12578_wink.ps}
\epsfbox{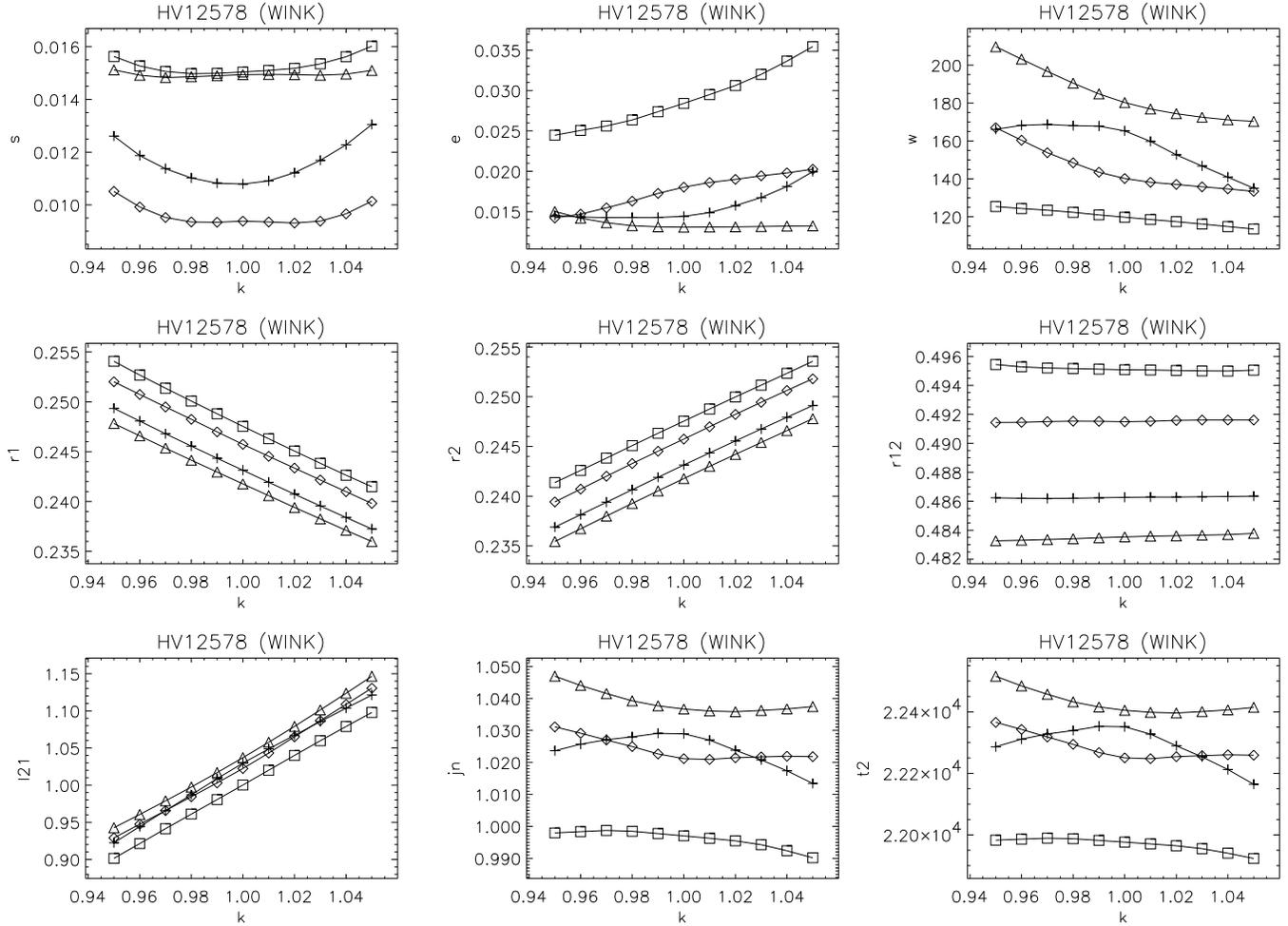}
\caption[]{\label{fig:hv12578_wink}
Results from WINK analyses of  HV12578 
for an adopted orbital inclination $i = 90\fdg0$, and 
$k = r_s/r_p$ values between 0.95 and 1.05. 
The individual figures show
s:  r.m.s. error (mag),
e:  orbital eccentricity $e$, 
w:  longitude of periastron $\omega$ ($\degr$),
r1: relative radius for primary component $r_p$,
r2: relative radius for secondary component $r_s$,
r12: sum of relative radii,
l21: luminosity ratio $L_s/L_p$, 
jn:  surface flux ratio,
t2:  effective temperature of secondary component $T_s$ (K).
Symbols are: 
cross $y$, 
diamond $b$, 
triangle $v$, 
square $u$.
}
\end{figure*}

\begin{figure*} % figure 14
\epsfxsize=185mm
%\epsfbox{hv12578_res.ps}
\epsfbox{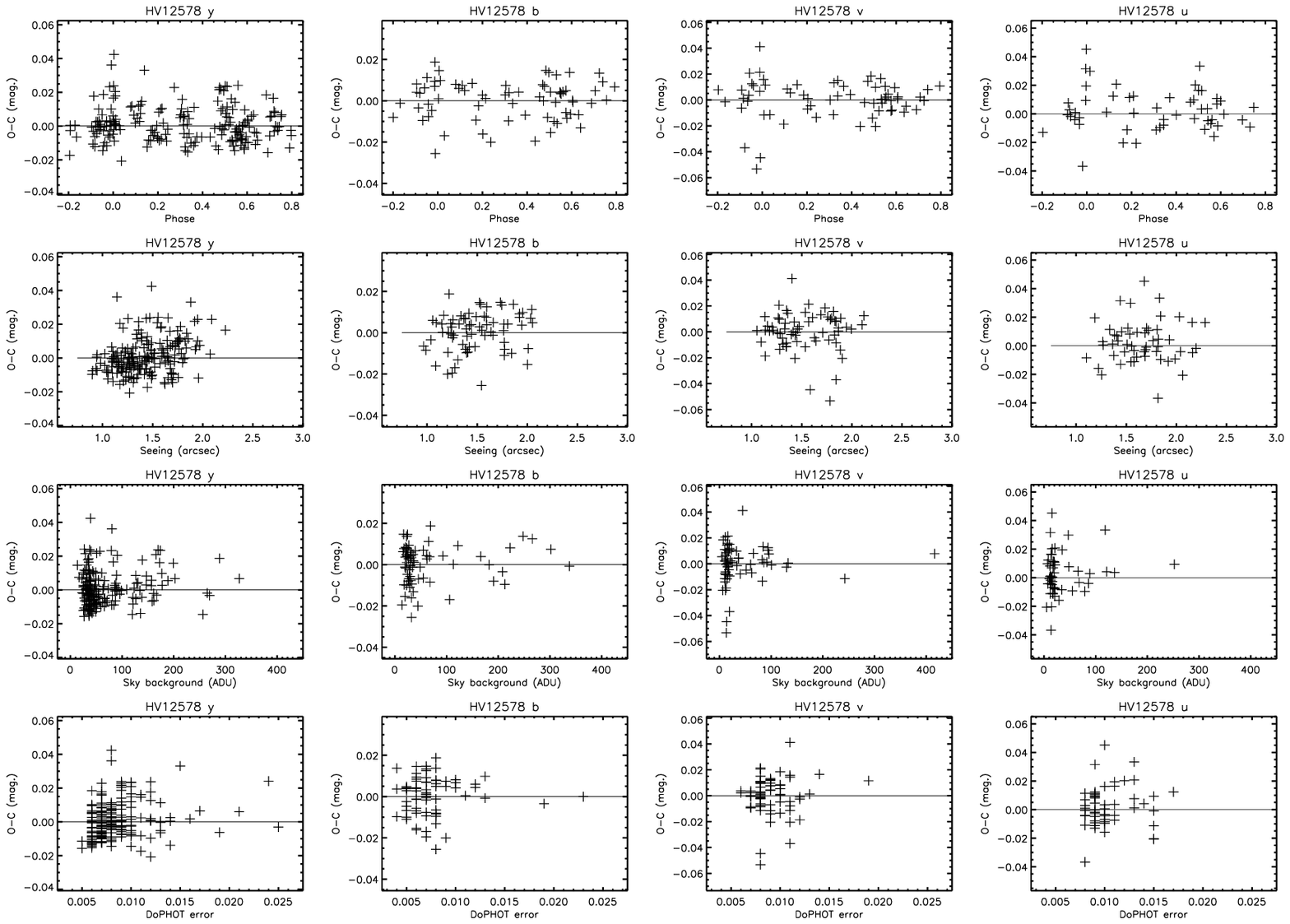}
\caption[]{\label{fig:hv12578_res}
$O-C$ residuals for the WINK solutions of HV12578 for  $i = 90 \degr$ and $k = 1.00$ 
given in Table \ref{tab:hv12578_wink} plotted versus orbital phase, seeing,
sky background, and DoPHOT error (mag).}
\end{figure*}

Since $i = 90\fdg$ formally gives the best fit we adopt this value, 
and in Fig. \ref{fig:hv12578_wink} solutions from all four light curves are shown. 
The $y$ and $b$ results agree well, but due to the incomplete phase 
coverage in $v$ and $u$, the elements from these two bands are less reliable. 
For the final solutions, we have 
chosen for all bands to fix $e$ and $\omega$ at the $y$ results. Such solution for
$k$ = 0.98, 1.00, and 1.02  are presented in Table \ref{tab:hv12578_wink}, 
and the $O-C$ residuals for the $k = 1.00$ solutions are shown in Fig. \ref{fig:hv12578_res}.
The $uvby$ solutions agree quite well, and no systematic trends exist in
the residuals. In all four bands, the component eclipsed near phase
0.49 and by us referred to as the secondary, is slightly more luminous than 
the (primary) component eclipsed at phase 0.00.

\subsubsection{Wilson-Devinney analyses}
\label{sec:hv12578_wd}

For comparison, a few WD solutions
for the $y$ light curve, done for very nearly the same ratio of radii $k$ as
the WINK solutions listed in Table \ref{tab:hv12578_wink}, 
are given in Table \ref{tab:hv12578_wdy}.
Taking into account the effect from the different treatment of the 
expansion effects on the resulting radii, explained in Sect.~\ref{sec:hv982_wd},  
we find only small differences between the WINK and WD results.
We notice a slightly higher surface flux ratio (via $T_s$) for the WD solution.

\subsubsection{Adopted photometric elements}
\label{sec:hv12578_phel}

The adopted mean photometric elements are listed in Table~\ref{tab:hv12578_phel}.
The components are very nearly identical, and as
seen, their relative sizes are slightly larger than for HV982. 
Relative radii accurate to 1\% have been established, together with well
established elements for the very slightly eccentric orbit.
Thus, when the results from the analysis of the VLT/UVES spectra are available,
we expect to be able to present accurate astrophysical parameters and distance
for HV12578.

\section{HV1433 (SMC)}
\label{sec:hv1433}
%SAO9, Table 
%SAO9 (jvc) 2424431.682  2.046987

\begin{figure*} % figure 15
\epsfxsize=185mm
%\epsfbox{hv1433teo.ps}
\epsfbox{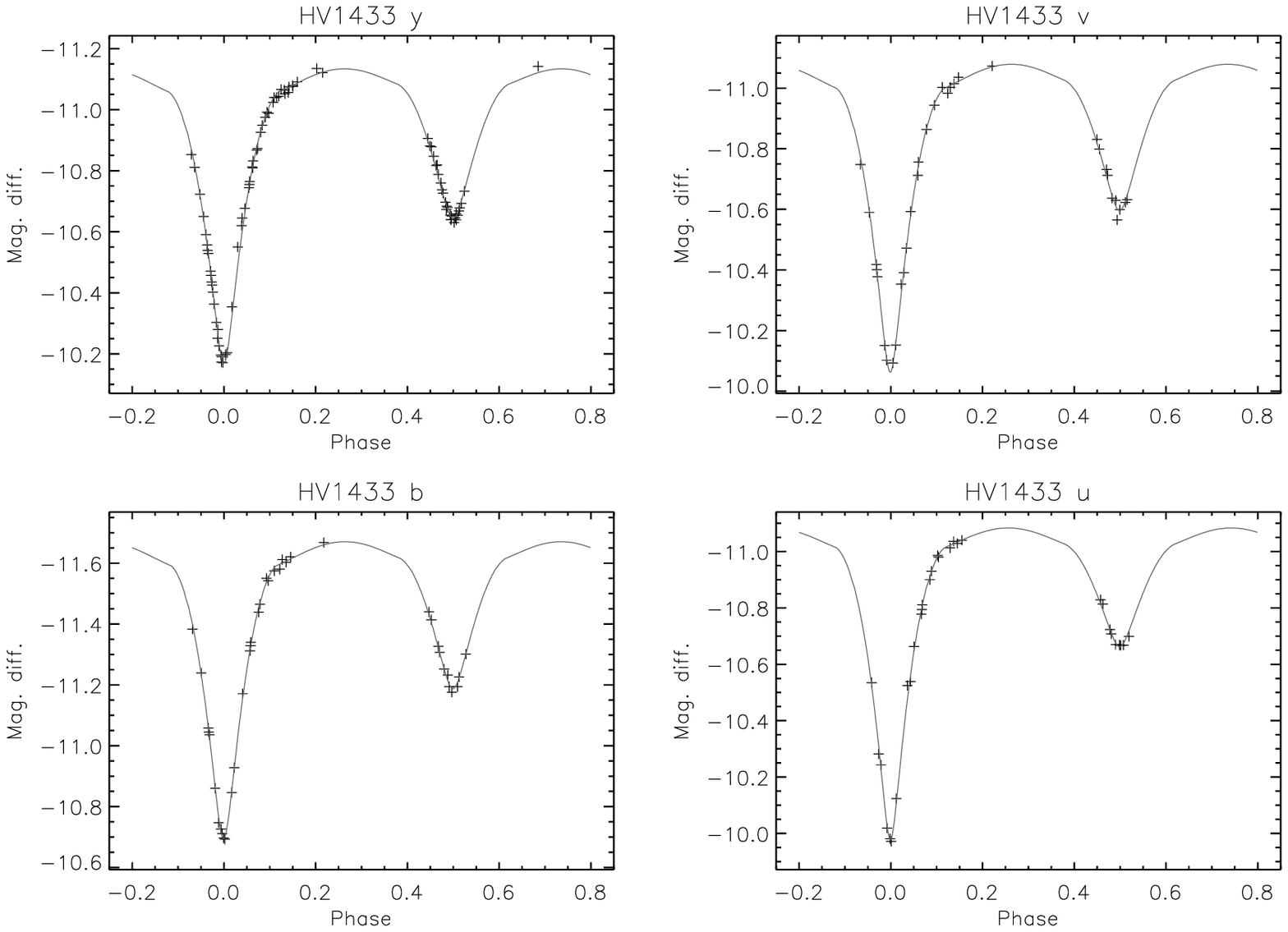}
\caption[]{\label{fig:hv1433}
$uvby$ light curves of HV1433 (SMC).
The theoretical curves correspond to the solutions ($q = 0.50$) given in Table \ref{tab:hv1433_wd5}.
}
\end{figure*}

At the time of the $uvby$ observations, only about 30 SMC eclipsing 
binaries were known, and the selection of candidates therefore more
difficult than for LMC. Also, the quality of the photographic light
curves from the Harvard survey (Gaposhkin \cite{g77}) is rather low,
and it was not possible to identify any clearly well-detached systems.

CCD photometry has been obtained for several Harvard SMC systems 
(HV1620, Pritchard et al. \cite{jdpetal98};
{\object{HV1761}, Duncan et al. \cite{duncan93};
{\object{HV1876}, Jensen et al. \cite{jcg88};
{\object{HV2016}, Jensen et al. \cite{jcg88};
{\object{HV2208}, West et al. \cite{wetal92};
HV2226, Jensen et al. \cite{jcg88}, Bell et al. \cite{sbetal91}),
and we initially included four additional in our project, 
{\object{HV1346}, HV1433, {\object{HV1597}, and HV11284.
For HV1346 ($P = 66\fd9$) and HV1597 ($P = 4\fd2$), rather few observations
were obtained, partial light curves were obtained for HV1433,
and for HV11284 we obtained complete light curves. In the present
paper, we will only include the HV1433 and HV11284 observations.

As mentioned in Sect. \ref{sec:cand}, HV1433 
is located in the southern part of the central region of SMC, just south
of the ${\rm OGLE-II}$ ${\rm SMC\_SC4}$ field.
Except for a 1.8 mag ($y$) fainter star about 3\farcs7 east of HV1433
(see Fig. \ref{fig:hv1433_field}), which causes no problem for the PSF photometry,
the surrounding field is quite clean. 

Apart from the photographic Harvard light curve, no earlier observations
of HV1433 have been published.

\subsection{Light curves}
\label{sec:hv1433_lc}

HV1433 was observed on 19 nights between JD2448971 and JD2450040,
and  84($y$), 38 ($b$),  33 ($v$), and  30 ($u$) points were obtained
in the four bands. The light curves are given in Tables 33-36 and
shown in Fig. \ref{fig:hv1433}. 
Typical accuracies per point are given in Table \ref{tab:dophoterrors}.

Unfortunately, the coverage of the light curves is far from complete,
and it is problematic to evaluate from the light curve shapes if HV1433 
could maybe be detached, or if it is a semi-detached system, possibly 
with the secondary component filling its Roche lobe, like e.g. HV2208 and HV2226.
However, the latter appears to be the most likely case.
As the depth of secondary eclipse is only about half of that of primary,
the temperatures of the components must be quite different.
The orbit appears to be circular.
In Sect. \ref{sec:hv1433_lcan} we will sketch possible photometric
solutions.

\subsection{Ephemeris}
\label{sec:hv1433_eph}

\begin{table}
\caption[]{\label{tab:hv1433_tmin}
Times of minima for HV1433 (LMC). 
Phases  are calculated from the ephemeris given in Eq. \ref{eq:hv1433_eph}.
References are: SAO9, Payne-Gaposchkin \& Gaposchkin (\cite{pgg66});
C2003, this paper.
}
\begin{flushleft}
\begin{tabular}{llcrc} \hline
\hline
HJD-2400000  & r.m.s.    & Type &   Phase    &  Reference  \\    
\hline
24431.682  &         &P&  0.0000&  SAO9  \\ 
50037.6300 &   0.0050&P&  0.0000&  C2003 \\
50038.6544 &   0.0005&S&  0.5004&  --\\
50040.7055 &   0.0010&S&  0.5024&  --\\
\hline
\end{tabular}
\end{flushleft}
\end{table}
 % HV1433 times of minima, table 16

One time of primary minimum and two of secondary minimum, derived
from the $uvby$ CCD observations, are given in Table~\ref{tab:hv1433_tmin}
together with the epoch of the ephemeris by 
Payne-Gaposchkin \& Gaposchkin (\cite{pgg66}).
We adopt the following ephemeris, derived from the new time of
primary minimum and the old epoch

\begin{equation}
\label{eq:hv1433_eph}
\begin{tabular}{r r c r r}
{\rm Min \, I} =  & 2450037.630 & + & $2\fd 047002$ &$\times\; E$ \\
                  &      $\pm 5$&   &       $\pm 5$ &             \\
\end{tabular}
\end{equation}

The period is slightly larger than that derived from the photographic
Harvard observations (2\fd 046987).

\subsection{Light curve analysis, photometric elements}
\label{sec:hv1433_lcan}

\begin{table*}
\caption[]{\label{tab:hv1433_wd5} 
WD solutions for HV1433 for adopted mass ratios between 0.4 and 0.6.
A semi-detached configuration is assumed, and $T_p = 23000$ K has been adopted throughout.
}
\begin{flushleft}
\begin{tabular}{lrrrrrr} \hline
\hline
                   &  $y$    &  $y$    &   $b$   &  $v$    &   $u$   & $y$\\
\hline
$q = m_s/m_p$      &  0.40   &  0.50   &  0.50   &  0.50   & 0.50    & 0.60\\

$i (\degr)$        &  88.37  &  86.17  &  84.43  & 86.15   & 86.68   & 84.80\vspace{-0.8mm}\\
                   & $\pm20$ & $\pm15$ & $\pm18$ & $\pm37$ & $\pm44$ & $\pm10$\\

$\Omega_p$         & 3.3394  & 3.4093  & 3.5562  & 3.4710  & 3.4474  & 3.6630\vspace{-0.8mm}\\
                   & $\pm84$ & $\pm192$& $\pm398$& $\pm557$& $\pm425$& $\pm201$\\

$\Omega_s$         & 2.6781  & 2.8758  & 2.8758  & 2.8758  & 2.8758  & 3.0624\\

$r_p$              &  0.347 & 0.352  &  0.334    & 0.344   & 0.347   & 0.333\\
$r_s$              &  0.303 & 0.321  &  0.321    & 0.321   & 0.321   & 0.336\\

$x_p$              &  0.34   & 0.34  &   0.38    & 0.40    & 0.40    & 0.34\\
$x_s$              &  0.37   & 0.37  &   0.43    & 0.46    & 0.43    & 0.37\\

$T_s$ (K)          &  16157  &  16150  & 16425   & 16577   & 16595   & 16180\vspace{-0.8mm}\\
                   &$\pm 67$ &$\pm 54$ &$\pm 70$ & $\pm102$& $\pm117$& $\pm 51$\\

$L_s/L_p$          &  0.455 &   0.494  &  0.539  & 0.494   & 0.356   & 0.600\\

$\sigma ({\rm mag}$)& 0.0134&   0.0111 &  0.0098 & 0.0148  & 0.0150  & 0.0120\\

\hline
\end{tabular}
\end{flushleft}
\end{table*}
 % HV1433 WD semidetached solutions, table 17

We have applied only the Wilson-Devinney model for the photometric analysis of
HV1433, since its components are expected to 
be larger and more deformed than appropriate for the WINK model; 
see Sect. \ref{sec:hv982_lcan} for details on the actual WD code.
A semi-detached configuration with the secondary filling its
Roche lobe has been assumed, since initial tests showed that detached
models converges towards this configuration.

The standard $uvby$ indices for the combined light of the components,
listed in Table~\ref{tab:std0}, correspond to an effective temperature of
about 23000 K (Napiwotzki et al. \cite{nap92}), which has then been adopted
for the primary component, although it might be slightly on the low side due
to the effect on the indices of the cooler component. Linear limb 
dar\-ke\-ning coefficients by Diaz-Cordoves et al. (\cite{dcg95}) were adopted 
together with gravity exponents and reflection albedos of 1.0. A circular orbit 
and synchronized rotation of the components were assumed.

Solutions for a range of fixed mass ratios indicate that values
below 0.4 and above 0.6 are not very likely. In Table~\ref{tab:hv1433_wd5}
we present a few solutions within the remaining range for which 
acceptable fits are reached.
Definitive solutions can not be obtained from the incomplete light curves,
but relative radii of around 0.34 (primary) and 0.32 (secondary), a 
temperature difference between the components of about 7000 K, and an
orbital inclination of about 86\degr are estimated.

\subsection{Discussion}
\label{sec:hv1433:disc}

Due to its brightness, its position within the SMC, and its
uncomplicated surrounding field, HV1433 is an interesting target
for future photometric and spectroscopic observations and studies, 
primarily for distance determinations. 
On the other hand, many of the new eclipsing binaries in the central region 
of SMC, now known from the ${\rm OGLE-II}$ project, could just as well
be considered. They include about 150 detached systems, most of which
are, however, much fainter than HV1433 and therefore in general more
difficult to observe accurately enough.

\section{HV11284 (SMC)}
\label{sec:hv11284}
%SAO9 (jvc) 2434682.321  3.625487   but shifted

\begin{figure*} % figure 16
\epsfxsize=185mm
%\epsfbox{hv11284teo.ps}
\epsfbox{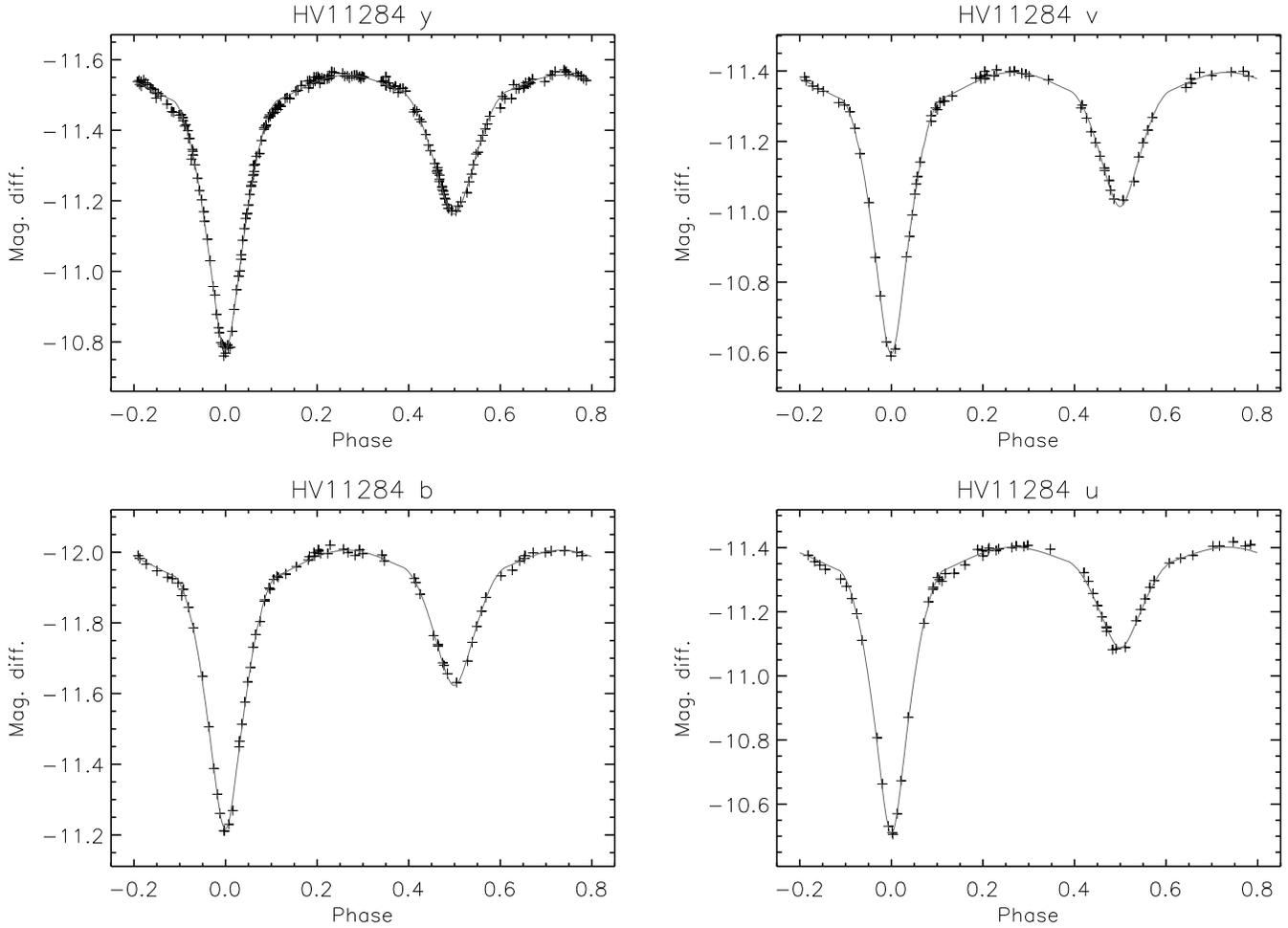}
\caption[]{\label{fig:hv11284}
$uvby$ light curves of HV11284 (SMC).
The theoretical curves correspond to the solutions given in Table \ref{tab:hv11284_wd5}.
}
\end{figure*}

As mentioned in Sect.~\ref{sec:cand}, HV11284 
is located slightly west of the optical center of SMC in the 
${\rm OGLE-II}$  field ${\rm SMC\_SC5}$ and is
identical to the ${\rm OGLE-II}$ eclipsing binary ${\rm SMC\_SC5\_140701}$.
We selected HV11284 as one of the most promising systems among
the few Harvard candidates available and obtained complete CCD light
curves 1992-94. In addition, 
$BVI$ light curves from 1997-98 are now also available from 
${\rm OGLE-II}$ (Udalski et al. \cite{ogle98}).

Being nearer to the SMC center, the HV11284 field is more crowded than
that of HV1433 with a higher density of faint stars.
There are no bright, close companions to HV11284, but the shape of its
image on CCD exposures obtained at about 1 arcsec seeing indicates the
presence of a possible very close, but also quite faint companion
slightly to the east; see Fig.  \ref{fig:hv11284_field}. 
If real, it is included in the DoPHOT photometry of HV11284 and therefore may
add a small amount of 3rd light to the light curve data.
We estimate that it is at least 3-4 mag fainter than HV11284 at maximum light, 
leading to an upper level of about 6\% of possible 3rd light ($L_3/(L_p+L_s)$).

\subsection{Light curves}
\label{sec:hv11284_lc}

HV11284 was observed on 34 nights between JD2448971 and JD2449638,
and 219 ($y$), 81 ($b$), 70 ($v$), and 63 ($u$) points were obtained
in the four bands. The light curves are given in Tables 37-40 and
shown in Fig. \ref{fig:hv11284}. 
Typical accuracies per point are given in Table \ref{tab:dophoterrors}.

As seen from the relatively broad eclipses and the light variations
outside eclipses, HV11284 consists of two rather close and significantly
deformed components. It is, however, not quite obvious if the system is 
detached or semi-detached. 
The eclipses differ significantly in depths, and the secondary component 
eclipsed at phase 0.5 is clearly much cooler than the primary.
The orbit appears to be circular.

\subsection{Ephemeris}
\label{sec:hv11284_eph}

\begin{table}
\caption[]{\label{tab:hv11284_tmin}
Times of minima for HV11284 (LMC). 
Phases  are calculated from the ephemeris given in Eq. \ref{eq:hv11284_eph}.
References are: SAO9, Payne-Gaposchkin \& Gaposchkin (\cite{pgg66});
U1998, Udalski et al. (\cite{ogle98}) adjusted by $-0\fd0087$; C2003, this paper.
}
\begin{flushleft}
\begin{tabular}{llcrc} \hline
\hline
HJD-2400000  & r.m.s.     & Type &   Phase    &  Reference  \\    
\hline
34682.321  &         &P&   0.0000 &  SAO9  \\ % shifted on plot ???
48975.607  &   0.005 &S&   0.4971 &  C2003 \\
49343.598  &   0.002 &P& $-0.0004$&  --\\
49633.636  &   0.002 &P&   0.0003 &  --\\
50471.1116 &         &P&   0.0001 &  U1998 \\   % shifted by -0.0087 d
\hline
\end{tabular}
\end{flushleft}
\end{table}
 % HV11284 times of minima, table 18

Two times of primary minimum and one of secondary minimum, derived
from the $uvby$ CCD observations, are given in Table~\ref{tab:hv11284_tmin}
together with the epochs of the ephemeris by 
Payne-Gaposchkin \& Gaposchkin (\cite{pgg66}) and Udalski et al. (\cite{ogle98}).
Based on a reanalysis of the OGLE $I$ light curve, the latter has been adjusted by $-0\fd0087$.
We note that the light curve tabulated by Gaposhkin (\cite{g77}) is shifted with 
primary minimum located at phase 0.25, i.e. as if the epoch should correspond
to maximum light level.
From a weighted least squares fit to the times of primary eclipse and the two epoch, 
we derive the following linear ephemeris:

\begin{equation}
\label{eq:hv11284_eph}
\begin{tabular}{r r c r r}
{\rm Min \, I} =  & 2449343.5996 & + & $3\fd 6254398$ &$\times\; E$ \\
                  &      $\pm  8$&   &      $\pm   4$ &             \\
\end{tabular}
\end{equation}

The period is somewhat shorter than that determined from the photographic 
Harvard observations (3\fd 625487), and somewhat longer than the result (3\fd62537)
published by Udalski et al. (\cite{ogle98}).
However, a reanalysis of their I (DIA; see below) light curve yields 3\fd62545.

\subsection{Light curve analysis, photometric elements}
\label{sec:hv11284_lcan}

\begin{table}
\caption[]{\label{tab:hv11284_ogle} 
WD solutions for HV11284 based on the two available OGLE $I$ light curves;
see text for details.
A mass ratio $q = 0.80$ and $T_p = 23000$ K has been adopted.
}
\begin{flushleft}
\begin{tabular}{lrr} \hline
\hline
            &  $I(DIA)$&  $I$     \\
\hline
$i (\degr)$ & 84.32    &  84.66\vspace{-0.8mm}\\
            &$\pm14$   & $\pm44$\\

$\Omega_p$  & 3.6172   & 3.6325\vspace{-0.8mm}\\
           &$\pm  49$  & $\pm138$\\

$\Omega_s$  & 4.1145   & 4.0562\vspace{-0.8mm}\\
            &$\pm 97$  &$\pm372$\\

$r_p$       & 0.367    &  0.365 \\
$r_s$       & 0.270    &  0.275  \\

$x_p$       & 0.24     &  0.24\\
$x_s$       & 0.31     &  0.31\\

$T_s$ (K) & 15242      &  15183\vspace{-0.8mm}\\
          &$\pm 49$    &$\pm 95$ \\

$L_s/L_p$ & 0.295      &  0.309\\

$\sigma ({\rm mag}$)&  0.0090& 0.0171\\
\hline
\end{tabular}
\end{flushleft}
\end{table}
 % HV11284 WD OGLE two I solutions q=0.80, table 19
%\input hv11284_wdy.tex % HV11284 WD y solutions for k=0.6-1.4, table 20
\begin{table}
\caption[]{\label{tab:hv11284_wdy} 
WD solutions for HV11284 ($y$ light curve) for adopted mass ratios between 0.6 and 1.4.
A detached configuration is assumed, and $T_p = 23000$ K has been adopted throughout.
}
\begin{flushleft}
\begin{tabular}{lrrrrr} \hline
\hline
$q = m_s/m_p$      &  0.60   &  0.80   &  1.00   &  1.20   & 1.40\\
\hline
$i (\degr)$        &  84.78  &  84.95  &  85.37  &  84.38  &  82.75\vspace{-0.8mm}\\
                   & $\pm11$ & $\pm10$ & $\pm 8$ & $\pm10$ & $\pm 7$\\

$\Omega_p$         & 3.3971  & 3.6418  & 3.9088  & 4.1889  & 4.4898\vspace{-0.8mm}\\
                   & $\pm61$ & $\pm49$ & $\pm30$ & $\pm39$ & $\pm42$\\

$\Omega_s$         & 3.3725  & 4.0714  & 4.8617  & 5.5874  & 5.9983\vspace{-0.8mm}\\
                   &$\pm59$  & $\pm101$& $\pm 94$& $\pm148$& $\pm175$\\

$r_p$              &  0.3679 & 0.3633  & 0.3557  & 0.3466  & 0.3353\\
$r_s$              &  0.2824 & 0.2737  & 0.2622  & 0.2588  & 0.2726\\

$F_p$              &  0.91   & 0.94    &  0.91   &  0.97   &  0.95 \\
$F_s$              &  0.84   & 0.82    &  0.74   &  0.72   &  0.72 \\

$x_p$              &  0.34   & 0.34    & 0.34    & 0.34    & 0.34\\
$x_s$              &  0.37   & 0.37    & 0.37    & 0.37    & 0.37\\

$T_s$ (K)          &  15160  &  14790  & 14393   &  14212  &  14249\vspace{-0.8mm}\\
                   &$\pm 40$ &$\pm 37$ &$\pm 35$&$\pm 34$ &$\pm 34$\\

$L_s/L_p$          &  0.314 &   0.286  &  0.260  &  0.259  & 0.304\\

$\sigma ({\rm mag}$)& 0.0097&   0.0086 &  0.0080 &  0.0078 & 0.0079\\

\hline
\end{tabular}
\end{flushleft}
\end{table}
 % HV11284 WD y solutions for k=0.6-1.4, table 20
%\input hv11284_wdi.tex % HV11284 WD I solutions for k=0.6-1.4, table 21
\begin{table}
\caption[]{\label{tab:hv11284_wdi} 
WD solutions for HV11284 (OGLE $I$ (DIA) light curve) for adopted mass ratios between 0.6 and 1.4.
A detached configuration is assumed, and $T_p = 23000$ K has been adopted throughout.
}
\begin{flushleft}
\begin{tabular}{lrrrrr} \hline
\hline
$q = m_s/m_p$      &  0.60   &  0.80   &  1.00   &  1.20   & 1.40\\
\hline
$i (\degr)$        &  83.22  &  84.32  &  84.18  &  83.29  &  81.86\vspace{-0.8mm}\\
                   & $\pm23$ & $\pm14$ & $\pm14$ & $\pm 9$ & $\pm 7$\\

$\Omega_p$         & 3.3861  & 3.6172  & 3.8983  & 4.1957  & 4.5005\vspace{-0.8mm}\\
                   & $\pm97$ & $\pm51$ & $\pm47$ & $\pm35$ & $\pm41$\\

$\Omega_s$         & 3.3861  & 4.1114  & 4.8585  & 5.4805  & 5.8599\vspace{-0.8mm}\\
                   &$\pm142$ & $\pm 97$& $\pm184$& $\pm128$& $\pm122$\\

$r_p$              &  0.3695 & 0.3668  & 0.3572  & 0.3457  & 0.3340\\
$r_s$              &  0.2868 & 0.2696  & 0.2625  & 0.2651  & 0.2803\\

$x_p$              &  0.24   & 0.24    & 0.24    & 0.24    & 0.24\\
$x_s$              &  0.31   & 0.31    & 0.31    & 0.31    & 0.31\\

$T_s$ (K)          &  15648  &  15242  & 14999   &  14870  &  14985\vspace{-0.8mm}\\
                   &$\pm 58$ &$\pm 49$ &$\pm 45$&$\pm 43$ &$\pm 43$\\

$L_s/L_p$          &  0.349 &   0.294  &  0.279  &  0.302  & 0.364\\

$\sigma ({\rm mag}$)& 0.0100&   0.0090 &  0.0080 &  0.0075 & 0.0075\\

\hline
\end{tabular}
\end{flushleft}
\end{table}
 % HV11284 WD I solutions for k=0.6-1.4, table 21
%\input hv11284_wdy3.tex % HV11284 WD y solutions with 3rd lt, table 22
\begin{table}
\caption[]{\label{tab:hv11284_wdy3} 
WD solutions for HV11284 ($y$ light curve) for adopted mass ratios between 0.6 and 1.4.
A detached configuration has been assumed, and $T_p = 23000$ K has been adopted throughout.
A small amount of 3rd light of 0.06 ($L_3/(L_p+L_s))$ has been included.
}
\begin{flushleft}
\begin{tabular}{lrrrrr} \hline
\hline
$q = m_s/m_p$      &  0.60   &  0.80   &  1.00   &  1.20   & 1.40\\
\hline
$i (\degr)$        &  86.02  &  86.04  &  86.06  &  86.29  &  84.83\vspace{-0.8mm}\\
                   & $\pm13$ & $\pm13$ & $\pm10$ & $\pm 8$ & $\pm 8$\\

$\Omega_p$         & 3.4103  & 3.6442  & 3.9034  & 4.1715  & 4.4576\vspace{-0.8mm}\\
                   & $\pm47$ & $\pm45$ & $\pm35$ & $\pm34$ & $\pm40$\\

$\Omega_s$         & 3.2994  & 3.9514  & 4.6699  & 5.4515  & 6.0543\vspace{-0.8mm}\\
                   &$\pm42$  & $\pm83$ & $\pm 80$& $\pm96$ & $\pm 93$\\

$r_p$              &  0.3659 & 0.3629  & 0.3565  & 0.3490  & 0.3394\\
$r_s$              &  0.2931 & 0.2859  & 0.2766  & 0.2669  & 0.2696\\

$x_p$              &  0.34   & 0.34    & 0.34    & 0.34    & 0.34\\
$x_s$              &  0.37   & 0.37    & 0.37    & 0.37    & 0.37\\

$T_s$              &  15280  &  14934  & 14595   &  14307  &  14175\vspace{-0.8mm}\\
                   &$\pm 38$ &$\pm 35$ &$\pm 33$&$\pm 33$ &$\pm 34$\\

$L_s/L_p$          &  0.347 &   0.318  &  0.293  &  0.273  & 0.288\\

$\sigma ({\rm mag}$)& 0.0095&   0.0085 &  0.0077 &  0.0075 & 0.0079\\

\hline
\end{tabular}
\end{flushleft}
\end{table}
 % HV11284 WD y solutions with 3rd lt, table 22
%\input hv11284_wd4.tex % HV11284 WD uvby solutions q=0.8, table 23
\begin{table}
\caption[]{\label{tab:hv11284_wd4} 
WD solutions for HV11284 for an adopted mass ratio of $q = 0.8$.
A detached configuration has been assumed, 
and $T_p = 23000$ K has been adopted throughout.
}
\begin{flushleft}
\begin{tabular}{lrrrrr} \hline
\hline
            &  $y$    &  $b$    &  $v$    &  $u$    \\
\hline
$i (\degr)$ &  84.95  &  85.00  &  85.10  &  84.52\vspace{-0.8mm}\\
            & $\pm10$ & $\pm20$ & $\pm32$ & $\pm45$\\

$\Omega_p$  & 3.6418  & 3.6849  & 3.7161  & 3.7443\vspace{-0.8mm}\\
            & $\pm49$ & $\pm76$ & $\pm105$& $\pm162$\\

$\Omega_s$  & 4.0714  & 4.1212  & 4.1270  & 3.9734\vspace{-0.8mm}\\
            &$\pm101$ & $\pm164$& $\pm288$& $\pm479$\\

$r_p$       &  0.363  & 0.357   & 0.353   & 0.349 \\
$r_s$       &  0.274  & 0.269   & 0.268   & 0.284 \\

$x_p$       &  0.34   & 0.38    & 0.40    & 0.40\\
$x_s$       &  0.37   & 0.43    & 0.46    & 0.43\\

$T_s$ (K)   &  14790  &  14852  & 15310   &  14178\vspace{-0.8mm}\\
            &$\pm 37$ &$\pm 65$ &$\pm 72$&$\pm 79$\\

$L_s/L_p$   &  0.286 &   0.272  &  0.275  &  0.212\\

$\sigma ({\rm mag}$)&0.0086&   0.0086 &  0.0084 &  0.0101\\
\hline
\end{tabular}
\end{flushleft}
\end{table}
 % HV11284 WD uvby solutions q=0.8, table 23
%\input hv11284_wd5.tex % HV11284 WD I,uvby solutions q=0.4 semidetached, table 24
\begin{table}
\caption[]{\label{tab:hv11284_wd5} 
WD solutions for HV11284 for an adopted mass ratios of $q = 0.4$.
A semidetached configuration with the less massive secondary component
filling its Roche lobe has been assumed, and $T_p = 23000$ K has been adopted 
throughout.
}
\begin{flushleft}
\begin{tabular}{lrrrrrr} \hline
\hline
            &  $I(DIA)$&  $y$    &  $b$    &  $v$    &  $u$    \\
\hline
$i (\degr)$ &  81.77   &  82.71  &  82.26  &  82.10  &  82.97\vspace{-0.8mm}\\
            & $\pm 6$  & $\pm 5$ & $\pm 8$ & $\pm 8$ & $\pm16$\\

$\Omega_p$  &  3.4145  & 3.3410  & 3.4436  & 3.7161  & 3.3378\vspace{-0.8mm}\\
            & $\pm142$ & $\pm81$ & $\pm110$& $\pm105$& $\pm253$\\

$\Omega_s$  &  2.6781  & 2.6781  & 2.6781  & 2.6781  & 2.6781\\

$r_p$       &  0.338   &  0.347  & 0.334   & 0.324   & 0.347 \\
$r_s$       &  0.303   &  0.303  & 0.303   & 0.303   & 0.303 \\

$x_p$       &  0.24    &  0.34   & 0.38    & 0.40    & 0.40\\
$x_s$       &  0.31    &  0.37   & 0.43    & 0.46    & 0.43\\

$T_s$ (K)   &  16056   &  15638  &  15773  & 16175   &  15166\vspace{-0.8mm}\\
            & $\pm 43$ &$\pm 37$ &$\pm 85$ &$\pm 61$&$\pm 70$\\

$L_s/L_p$   &  0.506   &  0.446 &   0.462  &  0.486  &  0.305\\

$\sigma ({\rm mag}$)& 0.0081  &0.0092&   0.0085 &  0.0086 &  0.0119\\
\hline
\end{tabular}
\end{flushleft}
\end{table}
 % HV11284 WD I,uvby solutions q=0.4 semidetached, table 24

From the shape of the light curves, the two components of HV11284 are expected to
be larger and more deformed than appropriate for the WINK model, and we have 
therefore applied only the Wilson-Devinney model for the photometric analysis; 
see Sect. \ref{sec:hv982_lcan} for details on the actual WD code.
The standard $uvby$ indices for the combined light of the components,
listed in Table~\ref{tab:std0}, correspond to an effective temperature of
about 23000 K (Napiwotzki et al. \cite{nap92}), which has then been adopted
for the primary component. 
As for HV1433, the actual effective temperature might be slightly higher, 
but tests show that the possible difference is of no importance for the 
light curve analysis.
Linear limb darkening coefficients by Diaz-Cordoves
et al. (\cite{dcg95}) were adopted together with gravity exponents and
reflection albedos of 1.0. A circular orbit and synchronized rotation of the
components were assumed.

The stellar shapes depend on the mass ratio between the components, and on their
rotational velocities, which are unfortunately not known since spectroscopic
observations are not available. This means that we needed to explore various 
plausible configurations.

Besides the $uvby$ light curves, we have also analyzed the $I$ light curve
available from the OGLE WWW catalogue\footnote{http://sirius.astrouw.edu.pl}. 
$B$ and $V$ light curves
are also available, but they contain too few points for meaningful analyses.
Two versions of the light curves are available: The original light curves are
based on photometry derived from a modified DoPHOT package, and recently
new and more accurate light curves based on difference image analyses 
(DIA) have been published
(\.Zebru\'n, Soszy\'nski, \& Wo\'zniak \cite{zsw01}, 
\.Zebru\'n et al \cite{zetal01}). 
We have used the $I$ (DIA) light curve for the WD analyses.
It contains 312 points obtained through a large number of nights.
It is generally well co\-ve\-red, but rather few points are obtained during 
primary eclipse.
The original $I$ light curve leads to nearly identical but less well
defined WD solutions; see Table~\ref{tab:hv11284_ogle}.

First, we have considered HV11284 to consist of two detached components, and
a set of $y$ and $I$ solutions for a range of adopted mass ratios between 0.6 and 1.4
are listed in Table~\ref{tab:hv11284_wdy} and \ref{tab:hv11284_wdi}. For all mass ratios, 
the hotter primary component is larger than the 8-9000 K cooler se\-con\-dary component.
Filling factors $F$ (see e.g. Mochnacki \cite{mochnacki84}) are between 0.91 and 0.97 
for the primary, and between 0.84 and 0.72 for the secondary.
Formally, the best solutions correspond to a mass ratio of around 1.2, which is,
however, not very likely. For mass ratios above 1.0, and given the large
temperature difference, the more massive component is expected to have evolved
off the main sequence and to have become not only the cooler but also the larger
component. This is clearly not supported by the light curve solutions.
If, on the other hand, both components of HV11284 are still within
the main sequence band, current evolutionary models predict a mass ratio not
larger than about 0.6 for a primary temperature of around 23000 K and a 8-9000 K 
temperature difference between the components.
Attempts to improve the fit of the theoretical light curves to the observations
for mass ratios around 0.6 by adjusting rotational velocities, reflection coefficients,
and limb darkening coefficients (including using non-linear laws) have not been
very successful, and adding a small amount of third light from the possible
close companion does not help much either, as seen from Table~\ref{tab:hv11284_wdy3}. 
Without knowing the mass ratio, it is, however, still considered too early to 
exclude definitively that HV11284 is a detached system. 
Analyses of the $b$, $v$, and $u$ light curves lead to similar but less accurate 
results and conclusions.
In Table~\ref{tab:hv11284_wd4} we compare the $uvby$ solutions for an arbitrary
(although not very likely) mass ratio of 0.8 which, at the moment, seem to represent
the deformation of the components adequately, although not as well as 
higher mass ratios.

Next, we have considered HV11284 to be a semi-detached system with the cooler
component filling its Roche lobe. For this configuration, we also find it 
impossible to determine safely the mass ratio from the light curve solutions.
It is, however, seen that only values below 0.5-0.6 lead to acceptable agreement 
between the theoretical light curves and the observations. 
The quality of the fits is comparable to those for detached solutions, meaning
that a semi-detached configuration for HV11284 is certainly possible.
Solutions for an adopted mass ratio of $q = 0.4$ are listed in 
Table~\ref{tab:hv11284_wd5}. 

In conclusion, we find relative radii of 0.33-0.36 (primary) and 0.26-0.30
(secondary) for the components and a temperature difference of 
of 8-9000 K between them, but the actual configuration of HV11284 remains an 
open question. 
The orbital inclination is 82\degr-85\degr. 

\subsection{Discussion}
\label{sec:hv11284:disc}

The $uvby$ and OGLE $I$ light curves of HV11284 are well covered and of
quite high quality. They should allow definitive photometric 
solutions once the mass ratio is established from spectroscopic observations. 
HV11284 is potentially a valuable system for distance determination, and since
it is relatively bright, we suggest that such observations are done, e.g. at
the VLT.

\section{Concluding remarks}
\label{sec:concl}

With present-day facilities, precise studies of eclipsing binaries
in the LMC and the SMC have become possible. Such objects are excellent
distance indicators, and distance moduli better than $\pm 0.10$ mag can 
often be obtained. Also, accurate absolute dimensions for MC binary
components are important for the study of internal structure, mass
loss and evolution of massive metal-deficient stars.

We have in this paper, as part of a large scale project on MC eclipsing 
binaries, presented new complete $uvby$ light curves and analyses
of them for the LMC systems HV982 and HV12578, and the SMC system HV11284,
and partial results for the SMC system HV1433.
Accurate standard {\it uvby} indices have also been established
for each binary, and individual interstellar reddenings have been determined.

HV982 and HV12578 are well-detached systems with eccentric orbits, 
each consisting of components of comparable sizes, and accurate photometric elements 
have been established.
Adopting the spectroscopic elements given by Fitzpatrick et al. (\cite{fp02})
for HV982, we have derived absolute dimensions of its component which agree well 
with their results. A distance modulus of $V_0-M_V = 18.63 \pm 0.08$ is obtained, 
corresponding to a distance of $52.6 \pm 2.0$ kpc, which is in formal agreement 
with (although slightly larger than) their determination.
In a forthcoming paper, we will combine our photometric results with
those underway from high-dispersion spectra obtained with the UVES spectrograph 
on the ESO VLT and present absolute dimensions, chemical abundances and distances for
HV12578 and refined results for HV982.

HV1433 and HV11284 (identical to the ${\rm OGLE-II}$ system ${\rm SMC\_SC5\_140701}$)
both consist of two rather close, deformed and quite different stars.
Because the mass ratios between the components - and their rotation rates - are not 
known, definitive photometric elements could not be obtained. We have, however, 
presented a sample of possible photometric solutions.
Since HV11284 is re\-la\-ti\-ve\-ly bright and has complete $uvbyI$ light curves, 
we recommend that spectroscopy is obtained for this SMC system. 
In addition to spectroscopy, more photometry is also needed for HV1433.

\begin{acknowledgements}
This research, which is based on observations carried out at the Danish 
1.5m telescope at ESO, La Silla, Chile, has been supported by the Danish 
Natural Science Research Council
through research grants to the project "Structure and evolution of stars; 
the distance scale of the Universe. New insight from studies of eclipsing 
binaries and pulsating stars", carried out at Aarhus University and Copenhagen
University, and through its Centre for Ground-Based Observational 
Astronomy.
ESO is gratefully acknowledged for granting a 2-months studentship
to SSL in 1995, during which parts of the data reduction were carried out.
SSL was furthermore partially supported by National Science Foundation grant 
number AST9900732.
This research has made use of the Simbad database, operated
at CDS, Strasbourg, France.
\end{acknowledgements}

{}

\listofobjects

\begin{thebibliography}{}
\bibitem[1997]{macho97}                                % MACHO LMC
Alcock, C., Allsman, R. A., Alves, D. et al. 1997,
 AJ 114, 326
\bibitem[1984]{acn84}                                   % Dm Vir
Andersen, J., Clausen, J.V., Nordstr\"om, B. 1984,
A\&A 137, 281
\bibitem[1983]{ja83}                                    % QX Car
Andersen, J., Clausen, J. V., Nordstr\"om, B., \&
Reipurth, B. 1983, A\&A 121, 271
\bibitem[1991]{sbetal91}                               % HV2226 SMC  
Bell, S. A., Hill, G., Hilditch, R. et al. 1991, 
MNRAS 250, 119
%Bell, S. A., Hill, G., Hilditch, R. W., Clausen, J. V., 
%Reynolds, A.P., \& Gim/'enez, A. 1991, MNRAS 250, 119
\bibitem[1993]{sbetal93}                               % HV5936 LMC
Bell, S. A., Hill, G., Hilditch, R. W., Clausen, J. V., 
\& Reynolds, A. P. 1993, MNRAS 265, 1047
\bibitem[1993]{b93}                                    % ROMAPHOT
Buonnano, R., Buscema, G., Corsi, C. E., Ferraro, I.,
\& Iannicola, G. 1993, A\&AS 106, 275
\bibitem[2000]{jvc00}                                  % review article
Clausen, J. V. 2000, in {\it From Extrasolar Planets
to Cosmology: The VLT Opening Symposium}, ed. 
J. Bergeron \& A. Renzini, Springer, 225
\bibitem[1986]{jvc86}                                  % V451 Oph
Clausen, J. V., Gim\'enez, A., \& Scarfe, C. D.  1986, 
A\&A 167, 287
\bibitem[1997]{clausen97} 
Clausen, J. V., Larsen, S. S., Garcia, J. M.,             % uvby sec std
 Gim\'enez, A., \& Storm, J. 1997, A\&AS 122, 559
\bibitem[1978]{crawford78}                             % (b-y)_0 - c_0 relation
Crawford, D. L. 1978, AJ 83, 48
\bibitem[2000]{dekker00}                               % UVES
Dekker, H., D'Odorico, S., Kaufer, A. Delabre, B.,
\& Kotzlowski, H. 2000, in Proc. Conf. SPIE 4008-61
\bibitem[1995]{dcg95}                                  % limb darkening
Diaz-Cordoves, J., Claret, A., \& Gim\'enez, A.
1995, A\&AS 110, 329
\bibitem[2000]{dodorico00}                             % UVES
D'Odorico, S. 2000, The Messenger 99, 2
\bibitem[2000]{ferra00}
Ferrarese, L., Silbermann, N. A., Mould, J. R. et al. 
2000, PASP 112, 177
%Ferrarese, L., Silbermann, N. A., Mould, J. R., Stetson, P. B.,
%Saha, A., Freedman, W. L., \& Kennicutt, R. C. 2000, PASP 112, 177
\bibitem[1993]{duncan93}                               % HV1761
Duncan, S.P.R., Tobin, W., Watson, R.D., \&
Gilmore, A.C. 1993, MNRAS 265, 189
\bibitem[2002]{fp02}                                   % hv982 distance
Fitzpatrick, E.L., Ribas, I., Guinan, E.F. et al.
2002, ApJ 564, 260
%Fitzpatrick, E.L., Ribas, I., Guinan, E.F.,         
% DeWarf, L.E., Maloney, F.P., \& Mass, D.  2002, 
\bibitem[1996]{flower96}                               % BC
Flower, P. J. 1996, ApJ 469, 355
\bibitem[1970]{g70}                                    % LMC        
Gaposhkin, S. I. 1970, Spec. Rep. 310,
 Smithsonian Astrophys. Obs.
\bibitem[1977]{g77}                                    % SMC          
Gaposhkin, S. I. 1977, Spec. Rep. 380,
 Smithsonian Astrophys. Obs.
\bibitem[2000]{gibson2000}                             % LMC distance review
Gibson, B.K. 2000, 
Mem. Soc. Astron. Italiana, 71, 693                    %astro-ph/9910574
\bibitem[1983]{ggp83}                                  % apsidal motion
Gim\'enez, A., \& Garcia-Pelayo, J. 
1983, Ap\&SS 92, 203
\bibitem[1995]{gb95}                                   % apsidal motion
Gim\'enez, A., \& Bastero, M. 1995, Ap\&SS 226, 99
\bibitem[1995]{eros95}
Grison, P., Beaulieu, J.-P., Pritchard, J. D. et al.     % EROS LMC
 1995, A\&AS 74, 331
\bibitem[2001]{g01}                                    % HV2274
Groenewegen, M. A. T., \& Salaris, M. 2001, 
A\&A 366, 752
\bibitem[1998]{g98}                                    % HV2274
Guinan, E. F., Fitzpatrick, E. L., DeWarf, L. E.
et al. 1998, ApJ 509, L21
%Guinan, E. F., Fitzpatrick, E. L., DeWarf, L. E., 
% Malony, F. P., Maurone, P. A., Ribas, I., 
%Pritchard, J. D., Bradstreet, D. H., 
%\& G\'imenez, A. 1998, ApJ 509, L21
\bibitem[1967]{hw67}                                   % LMC atlas
Hodge, P.W., Wright, F.W. 1967,
The Large Magellanic Cloud (Washington)
\bibitem[1977]{hw77}                                   % SMC atlas
Hodge, P.W., Wright, F.W. 1977,
The Small Magellanic Cloud (Washington)
\bibitem[1988]{jcg88}                                  % CCD lc's
Jensen, K. S., Clausen, J. V., \& Gim\'enez, A. 1988,
A\&AS 74, 331
\bibitem[1992]{hksrf92}                                % MOMF
Kjeldsen, H.,\& Frandsen, S. 1992, PASP 104, 413
\bibitem[1972]{k72}
Kreiner, J. M. 1972, 
Rocz. Astr. Obs. Krakowskiego 43, 98
\bibitem[1992]{kurucz92}                               % uvby fluxes
Kurucz, R.K. 1992, in:
Barbuy B, Renzini A. (eds) Proc. IAU Symp. 149,
The Stellar Population of Galaxies, Reidel, Dordrecht, 
p.225
\bibitem[1956]{kvw56}                                  % tmin method
Kwee, K. K., \& van Woerden A. 1956, BAN 12, 327
\bibitem[1965]{lk65}                                   % eph method
Lafler, J., \& Kinman, T. D. 1965, ApJS 11,216
\bibitem[1999]{ssl99}                                  % YMC II, msynth
Larsen, S. S. 1999, A\&AS 139, 393
\bibitem[1996]{ssl96}                                  % ssl speciale
Larsen, S. S. 1996, Master Thesis,
Astronomical Observatory, Copenhagen University
\bibitem[2001]{lcs01}                                  % paper 1
Larsen, S. S., Clausen, J. V., \& Storm, J. 2001,             
A\&A, 364, 455 
\bibitem[1984]{mochnacki84}                            % Roche model paper
Mochnacki, S.W. 1984, ApJS 55,551
\bibitem[1992]{nap92}                                  % uvby Te
Napiwotzki, R., Sch\"onberner, D., \& Wenske V.
1993, A\&A 268, 653
\bibitem[2000]{n00}                                    % HV2274 phot
Nelsen, C. A., Cook, K. H., Popowski, P., 
\& Alves, D.R. 2000, AJ 119, 1205
\bibitem[1994]{nb94}                                   % HV2241, 1620
Niemela, V. S., \& Bassino, L. P. 1994, ApJ 437, 332
\bibitem[1990]{anlpv90}                                % WINK
Nordlund, \AA., \& Vaz, L. P. R. 1990, A\&A 228, 231
\bibitem[2000]{o00}                                    % HV2543 LMC
Ostrov, P. G., Lapasset, E., \& Morrell, N.I. 2000,
A\&A 356, 935
\bibitem[1997]{bp97}                                   % review article
Paczy\'{n}ski, B. 1997, in {\em The extragalactic 
distance scale}, STSI Symp. Ser.10, ed. M. Livio, 
M. Donahue \& N. Panagia, Cambridge Univ.  Press, 273
\bibitem[1971]{pg71}                                   % LMC   
Payne-Gaposchkin, C. H. 1971,
Smithson. Contrib. Astrophys., 13
\bibitem[1966]{pgg66}                                  % SMC
Payne-Gaposchkin, C. H., \& Gaposhkin, S. I. 1966,
Smithson. Contrib. Astrophys., 9
\bibitem[2000]{pn00}                                   % HV982 etc.     
Pritchard, J. D., \& Niemela, V. 2000, in 
{\it From Extrasolar Planets
to Cosmology: The VLT Opening Symposium}, ed. 
J. Bergeron \& A. Renzini, Springer, 232
\bibitem[1998a]{jdpetal98}                             % hv1620
Pritchard, J., Tobin, W., Clark, M., 
\& Guinan, E. F. 1998a, MNRAS 297, 278 
\bibitem[1998b]{jdp98b}                                % hv982 lc's
Pritchard, J., Tobin, W., Clark, M., 
\& Guinan, E.F. 1998b, MNRAS 299, 1087
\bibitem[2000]{ir00a}                                  % overshooting papaer
Ribas, I., Jordi, C., Gim\'enez, A., 2000a,
MNRAS 318, L55
\bibitem[2000]{ir00b}                                  % HV2274 abs dim
Ribas, I., Guinan, E. F., Fitzpatrick, E. L. et al.
2000b, ApJ 528, 692 
\bibitem[2002]{ir02}                                   % EROS 1044
Ribas, I., Fitzpatrick, E.L., Malony, F.P., \&        % 0204061  get final ref.
Guinan, E.F., 2002, ApJ 574, 771
%astro-ph\/0204061
\bibitem[2002]{sg02}                                   % B star distance method
Salaris, M., Groenewegen, M.A.T. 2002, A\&A 381,, 440              
\bibitem[1993]{schecter93}                             % DoPHOT
Schechter, P. L., Mateo, M., \& Saha, A. 1993,
PASP 105, 1342
\bibitem[1942]{sn42}
Shapley, H., Nail, V. McKibben 1942, 
Harv. Bull. 916, 19
\bibitem[1953]{sn53}
Shapley, H., Nail, V. McKibben 1953, 
Proc. Nat. Acad. Sci. U.S.A. 39, 1
\bibitem[1987]{stetson87}                              % DAOPHOT II 
Stetson, P. 1987 PASP 99, 191
\bibitem[1998a]{u98}                                   % HV2274 photometry
Udalski, A., Pietrzyn\'{s}ki, G., Wo\'{z}niak, P.
et al. 1998a, ApJ 509, L25
\bibitem[1998b]{ogle98}                                % OGLE SMC
Udalski, A., Soszy\'{n}ski, I., Szyma\'{n}ski, M. 
et al. 1998b, Acta Astron. 48, 563
\bibitem[1984]{lpv84}                                  % WINK
Vaz, L. P. R. 1984,  Ph.D. thesis, 
Copenhagen University (unpublished)
\bibitem[1986]{lpv86}                                  % WINK
Vaz, L. P. R. 1986, 
Rev. Mexicana Astron. Astrof. 12, 177
\bibitem[1995]{lpv95}                                  % WD extension (LZ Cen)
Vaz, L. P. R., Andersen, J., 
\& Rebello Soares, M. C. A. 1995, A\&A 301, 693
\bibitem[1985]{lpvan85}                                % WINK
Vaz, L. P. R., \& Nordlund, \AA. 1985, A\&A 147, 281
\bibitem[1992]{w92}
Watson, R. D., West, S. R. D., Tobin, W., \&           % HV2274 CCC lc 
Gilmore, A.C. 1992, MNRAS 258, 527
\bibitem[1993]{ws93}                                   % variability
Welch, D. L., Stetson, P. B. 1993, AJ 105, 1813
\bibitem[1992]{wetal92}                                % HV2208
West, S.R.D., Tobin, W., \& Gilmore, A.C. 1992,
MNRAS 254, 419
\bibitem[1990]{west1990} 
Westerlund, B. E. 1990 A\&AR 2, 29                      % Westerlund I
\bibitem[1997]{west1997} 
Westerlund, B. E. 1997, {\it The Magellanic Clouds},    % Westerlund II
Cambridge University Press
\bibitem[1994]{rew94}                                  % WD
Wilson, R. E. 1994, PASP 106, 921
\bibitem[1971]{rewd71}                                 % WD
Wilson, R. E., \& Devinney, E. J. 1971, ApJ 166, 605
\bibitem[1971]{dw71}                                   % WINK
Wood, D. B. 1971, AJ 76, 701
\bibitem[2001]{zsw01}                                  % DIA method
\.Zebru\'n, K., Soszy\'nski, I., and
Wo\'zniak, P.R. 2001, Acta Astron. 51, 303              %ph110612                    
\bibitem[2001]{zetal01}                                % DIA catalogue
\.Zebru\'n, K., Soszy\'nski, I., 
Wo\'zniak, P.R., et al. 2001, Acta Astron. 51, 317      %ph110623 
\end{thebibliography}
\end{document}